\documentclass[twocolumn]
{aastex631}
\let\splitbox\undefined
\usepackage{amsmath, amstext, natbib, verbatim, graphicx, xcolor, longtable, float
, hyperref, booktabs}
\usepackage[maxfloats=100]{morefloats}
\usepackage{adjustbox}

\newcommand{\OII}{\hbox{{\rm [O}\kern 0.1em{\sc ii}{\rm ]}}}
\newcommand{\NeIII}{\hbox{{\rm [Ne}\kern 0.1em{\sc iii}{\rm ]}}}
\newcommand{\OIII}{\hbox{{\rm [O}\kern 0.1em{\sc iii}{\rm ]}}}
\newcommand{\Hb}{\hbox{{\rm H}$\beta$}}
\newcommand{\Ha}{\hbox{{\rm H}$\alpha$}}
\newcommand{\NII}{\hbox{{\rm [N}\kern 0.1em{\sc ii}{\rm ]}}}
\newcommand{\SII}{\hbox{{\rm [S}\kern 0.1em{\sc ii}{\rm ]}}}

\newcommand{\HeI}{\hbox{{\rm He}\kern 0.1em{\sc i}}}
\newcommand{\HeII}{\hbox{{\rm He}\kern 0.1em{\sc ii}}}
\newcommand{\HII}{\hbox{{\rm H}\kern 0.1em{\sc ii}}}
\newcommand{\NeV}{\hbox{{\rm [Ne}\kern 0.1em{\sc v}{\rm ]}}}
\newcommand{\FeVII}{\hbox{{\rm [Fe}\kern 0.1em{\sc vii}{\rm ]}}}

\begin{document}

\title{\large \bf  
%The Most Extreme \Ha\ Emitters are Powered by AGN as Revealed by JWST}
OCEANS of Absorption: High-resolution NIRSpec Spectroscopy Reveals Diverse Balmer-line Absorption in Little Red Dots}
% AGN and Starburst Contribution to the Most Extreme Emission Lines at $4<z<9$} 

\shorttitle{OCEANS of Absorption }
\shortauthors{Davis et al.}

%%%
%Author Blocks

\correspondingauthor{Kelcey Davis}
\email{kelcey.davis@uconn.edu}

\author[0000-0001-8047-8351]{Kelcey Davis}
\altaffiliation{NSF Graduate Research Fellow}
\affiliation{Department of Physics, 196A Auditorium Road, Unit 3046, University of Connecticut, Storrs, CT 06269, USA}
\affil{Los Alamos National Laboratory, Los Alamos, NM 87545, USA}

\author[0000-0001-5384-3616]{Madisyn Brooks}
\altaffiliation{NSF Graduate Research Fellow}
\affiliation{Department of Physics, 196A Auditorium Road, Unit 3046, University of Connecticut, Storrs, CT 06269, USA}

\author[0000-0002-6386-7299]{Raymond C.\ Simons}
\affiliation{Department of Engineering and Physics, Providence College, 1 Cunningham Sq, Providence, RI 02918 USA}

\author[0000-0002-1410-0470]{Jonathan R. Trump}
\affiliation{Department of Physics, 196A Auditorium Road, Unit 3046, University of Connecticut, Storrs, CT 06269, USA}

\author[0000-0001-6813-875X]{Guillermo Barro}
\affiliation{Department of Physics, University of the Pacific, Stockton, CA 90340 USA}

%%%%%%%%%%%Alpha order%%%%%%%%%%%%%%%%%%%%%%%%%%%%%%%%%%%

\author[0000-0002-7959-8783]{Pablo Arrabal Haro}
\affiliation{Center for Space Sciences and Technology, UMBC, 5523 Research Park Dr, Baltimore, MD 21228 USA }
\affiliation{Astrophysics Science Division, NASA Goddard Space Flight Center, 8800 Greenbelt Rd, Greenbelt, MD 20771, USA}

\author[0000-0001-8534-7502]{Bren E. Backhaus}
\affil{Department of Physics and Astronomy, University of Kansas, Lawrence, KS 66045, USA}
% \email{bren.backhaus@ku.edu}

\author[0000-0001-7151-009X]{Nikko J. Cleri}
\affiliation{Department of Astronomy and Astrophysics, The Pennsylvania State University, University Park, PA 16802, USA}
\affiliation{Institute for Computational and Data Sciences, The Pennsylvania State University, University Park, PA 16802, USA}
\affiliation{Institute for Gravitation and the Cosmos, The Pennsylvania State University, University Park, PA 16802, USA}

\author[0000-0002-6219-5558, gname=Alexander, sname='de la Vega']{Alexander de la Vega}
\affiliation{Department of Physics and Astronomy, University of California, 900 University Ave, Riverside, CA 92521, USA}
%\email{alexandd@ucr.edu}

\author[0000-0001-8519-1130]{Steven L. Finkelstein}
\affiliation{Department of Astronomy, The University of Texas at Austin, Austin, TX, USA}
\affiliation{Cosmic Frontier Center, The University of Texas at Austin, Austin, TX, USA}
%\email{stevenf@astro.as.utexas.edu}

\author[0000-0002-7831-8751]{Mauro Giavalisco}
\affiliation{University of Massachusetts Amherst, 710 North Pleasant Street, Amherst, MA 01003-9305, USA}
%\email{mauro@umass.edu}

\author[0000-0001-9440-8872]{Norman A. Grogin}
\affiliation{Space Telescope Science Institute, 3700 San Martin Drive, Baltimore, MD 21218, USA}

\author[0000-0002-3301-3321]{Michaela Hirschmann}
\affiliation{Institute of Physics, Laboratory of Galaxy Evolution, Ecole Polytechnique Fédérale de Lausanne (EPFL), Observatoire de Sauverny, 1290 Versoix, Switzerland}

\author[0000-0001-6251-4988]{Taylor A. Hutchison}
\altaffiliation{NASA Postdoctoral Fellow}
\affiliation{Astrophysics Science Division, NASA Goddard Space Flight Center, 8800 Greenbelt Rd, Greenbelt, MD 20771, USA}

\author[0000-0002-8360-3880]{Dale Kocevski}
\affiliation{Department of Physics and Astronomy, Colby College, Waterville, ME 04901, USA}
%\email{dale.kocevski@colby.edu}

\author[0000-0002-6610-2048]{Anton M. Koekemoer}
\affiliation{Space Telescope Science Institute, 3700 San Martin Drive,
Baltimore, MD 21218, USA}

\author[0000-0000-0000-0001]{Erini Lambrides}
\affiliation{Astrophysics Science Division, NASA Goddard Space Flight Center, 8800 Greenbelt Rd, Greenbelt, MD 20771, USA}
\affiliation{Department of Astronomy, University of Maryland, College Park, MD 20742, USA}
\affiliation{Center for Research and Exploration in Space Science and Technology, NASA/GSFC, Greenbelt, MD 20771 USA}
%\email[show]{erini.lambrides@nasa.gov}

\author[0000-0003-1354-4296]{Mario Llerena}
\affiliation{INAF Osservatorio Astronomico di Roma, Via Frascati 33, 00078 Monte Porzio Catone, Rome, Italy}

\author[0000-0003-1581-7825]{Ray A. Lucas}
\affiliation{Space Telescope Science Institute, 3700 San Martin Drive, Baltimore, MD 21218, USA}
%\email{lucas@stsci.edu}

\author[0000-0001-6434-7845]{Madeline A. Marshall}
\affil{Los Alamos National Laboratory, Los Alamos, NM 87545, USA}

\author[0000-0001-8688-2443]{Elizabeth J.\ McGrath}
\affiliation{Department of Physics and Astronomy, Colby College, Waterville, ME 04901, USA}

\author[0000-0001-7503-8482]{Casey Papovich}
\affiliation{Department of Physics and Astronomy, Texas A\&M University, College Station, TX, 77843-4242 USA}
\affiliation{George P.\ and Cynthia Woods Mitchell Institute for Fundamental Physics and Astronomy, Texas A\&M University, College Station, TX, 77843-4242 USA}

\author[0009-0006-7692-245X]{Aidan Starrs}
\affiliation{Department of Engineering and Physics, Providence College, 1 Cunningham Sq, Providence, RI 02918 USA}

\author[0000-0003-1282-7454]{Anthony J. Taylor}
\affiliation{Department of Astronomy, The University of Texas at Austin, Austin, TX, USA}

\author[0000-0001-7968-3892]{Phoebe R. Upton Sanderbeck}
\affil{Los Alamos National Laboratory, Los Alamos, NM 87545, USA}

\author[0000-0002-9373-3865]{Xin Wang}
\affiliation{School of Astronomy and Space Science, University of Chinese Academy of Sciences (UCAS), Beijing 100049, China}
\affiliation{National Astronomical Observatories, Chinese Academy of Sciences, Beijing 100101, China}
\affiliation{Institute for Frontiers in Astronomy and Astrophysics, Beijing Normal University, Beijing 102206, China}

\author[0000-0003-3735-1931]{Stijn Wuyts}
\affiliation{Department of Physics, University of Bath, Claverton Down, Bath BA2 7AY, UK}

\begin{abstract}
The ``Little Red Dots'' (LRDs) that appeared in early JWST deep field images have been the subject of significant study since their discovery. In this work, we present high-resolution follow-up spectroscopy from the OCEANS program of 10 LRDs with \Ha\ coverage at $3<z<7$ in the CEERS/EGS legacy deep field. We find Balmer-line absorption in 4 of these LRDs, a $40\%$ detection rate, higher than the fractions reported in lower-resolution NIRSpec surveys. All of the absorbers are presented in high-resolution for the first time here and two have Balmer-line absorption detected for the first time. We demonstrate that this increased recovery of absorbers is partly tied to survey spectral resolution, with higher resolution recovering more absorbers.
Of the 10 LRDs, 7 are best fit by \Ha\ profiles with exponential wings. This includes 2 of the LRDs with absorption. 
%Four of the LRD absorbers are best fit by Gaussian profiles that fall off with exponential wings, consistent with either electron scattering or broad-line region stratification. Balmer-line absorption is compared to optical to UV and Balmer break colors. 
We find that absorbers tend to be blue-shifted with a median velocity offset of near-zero ($-49~\rm{km}~\rm{s}^{-1}$) and an absorption equivalent width of 5.3~\AA. 
%We showcase one of the bluest detected Balmer-absorbers as well as two low-luminosity LRDs with absorption and compare two LRDs with both \Ha\ and \Hb\  absorption. 
Trends are explored to compare LRD absorption velocity and equivalent width along the sequence of LRD properties. We find significant correlation between absorption velocity and LRD color and Balmer break strength. We confirm an LRD observed previously at medium-resolution has statistically significant absorption velocity offsets between \Ha\ and \Hb. The diversity of LRD absorption properties can be effectively explained by a model with a radial distribution of partial-covering absorbing gas that is often co-located near the broad-line emission regions, along with a radial gradient of close inflow and distant outflow velocities for the absorbing gas.
We present other interesting LRDs, including an outflow-dominated LRD (OCEANS 20504) and an LRD with a weak Balmer break and relatively blue UV-to-optical colors but clear Balmer-line absorption (OCEANS 102364).
This high occurrence of absorbing hydrogen in LRDs, evident by both the Balmer-line absorption features and Balmer break strengths, implies a near-ubiquitous presence of excited $n=2$ hydrogen near the central engines, with a diversity of absorption-line kinematics that likely represents a radial gradient of inflowing and outflowing gas. 

\end{abstract}

\section{Introduction}  
\label{sec:intro}

%A note on the introduction: Kelcey will be applying for post-docs soon and would like to try and land at a national lab. With this in mind, the introduction has been written in a way that physicists outside of astrophysics can read and understand. If this feels over-explainy or goes in to too much detail, please let me know. But the goal is that other physicsts can read this and understand what I am doing.

%Jon: intro- empiracal first THEN explanations: weird features, emission lines, DONT start with 
%physical interpretations/ BHs

The previous years of high-redshift ($z>4$) JWST discovery have been defined by the emergence of a new class of luminous sources in the early universe. %supermassive black holes (SMBHs) and the community convergence on a first model to explain their observed properties. These SMBHs were classified as  
These ``Little Red Dots'' \citep[LRDs, e.g:][]{Kocevski2023, Mathee2024, Kokorev2024} were unidentified by previous space-based telescopes but detected in almost all JWST extragalactic deep fields.  LRDs are defined by compactness, with angular sizes $\lesssim200$ pc, ``v-shaped'' spectral energy distributions (SEDs) with incongruous blue rest-UV and red rest-optical continuum slopes, and broad Balmer-lines \citep{Hviding2025}. %, and an SED
%kink associated with the ``Balmer break'' at $\lambda = 3646 $\AA. 

LRDs are abundant in the first $0.5$--$2$ billion years of cosmic time ($\sim4<z<9$) with UV and Bolometric number densities of $\Phi \sim 10^{-5}$-$10^{-4}\,\mathrm{cMpc^{-3}\,mag^{-1}}$ \citep{Kokorev2024}. At $z<4$, they become increasingly rare (see some exceptions in \citealt{Bisigello2025, Ji2026, Ma2026,  Rinaldi2025}). 
This over-abundance of atypical sources in the first few billion years implies that LRDs represent a phenomenon that is unique to the conditions of the early Universe. An explanation for these unexpected sources is key to understanding how early galaxies and black holes formed. %and co-evolved with supermassive black holes (SMBHs). %SMBHs grew and co-evolved with their host galaxies.

Balmer breaks, a prominent feature in most LRDs, are typical in nearby galaxies. They are signatures of excited (n=2) hydrogen common in the atmospheres of A type stars. In galaxies with such stars in abundance, hydrogen atoms have the correct energy to absorb rest-UV photons blueward of $\lambda = 3646 $\AA. This results in a forest of absorption features that blends together into a ``break'' in the SED. Such breaks in LRDs occur at their characteristic SED ``kink''. 
%While Balmer breaks in LRDs could be explained by the presence of old stellar populations within the LRD host galaxy, this would not explain the smooth slope present in the breaks of LRDs. A stellar break also fails to produce the strength of many LRD breaks, specifically in extreme cases \citep{Naidu2025}. 

Early work attempted to explain the Balmer breaks of LRDs with old stellar populations but found that the implied galaxy masses and ages were in conflict with the young age of the universe at high redshift \citep{Labbe2023, Boylan-Kolchin2023}. LRDs also have Balmer breaks that are much smoother and often much stronger than observed for stellar populations \citep{Naidu2025}. Together, this implies Balmer breaks in LRDs are driven by a process independent of, or in addition to, stellar light. Such a process must invoke a large density of hydrogen in the n=2 state to create the observed Balmer breaks.

%Such breaks were not associated with SMBHs prior to JWST. 
%However, 
Follow-up observations of LRDs indicate they may host accreting supermassive black holes (SMBHs) that dominate the rest-optical light output, as evidenced by their broadened Balmer lines \citep{Hviding2025, Greene2024, Mathee2024, Pang2026, Wang2025} and rest optical continuum slopes best fit by SMBH contribution  \citep{Kocevski2024}. 
A popular model that explains the LRD observed properties invokes, not star formation, but a SMBH veiled in a cocoon of dense, excited hydrogen gas \citep{Inayoshi2025, Torralba2025}. This is the only model thus far that can explain both the presence and strength of the Balmer breaks, including the extreme break-strength outliers, or ``cliff'' sources \citep{ deGraaffCLIFF2025, Naidu2025, Taylor2025}.

Recently, the physical mechanism that broadens the Balmer lines in these sources has been under question. In low-z quasars, most Balmer-line broadening is attributed to kinematics associated with rotating high density gas surrounding the SMBH known as the broad-line region (BLR). In many LRDs, the line-shape outside of the Gaussian core has an exponential profile, which requires stratification in the BLR \citep{Madau2026, Tang2026, Ji2026} or exponential scattering from dense hydrogen \citep{Rusakov2026}. This emission-line structure is not unique to LRDs and has been observed in many quasars identified by the Sloan Digital Sky Survey (SDSS) interpreted as evidence for thermal electron scattering \citep{Kollatschny2013, Laor2006}. Honing in on the mechanisms by which Balmer emission is broadened is another critical step in developing a complete model for LRDs.

% with suggestions that Balmer lines are broadened by electron scattering in the dense hydrogen shell. Still other works \citep{Madau2026, Tang2026, Ji2026} suggest that a stratified broad-line region may create the characteristic exponential shape in LRD Balmer lines. 

A subset of the LRDs also have narrow Balmer-line absorption features, frequently offset from the emission-line center \citep{Kocevski2024,Mathee2024,Matthee2026,Naidu2025,Taylor2024, Taylor2025, Lambrides2025}. Such absorption features should be present in all LRDs that have Balmer breaks if the break is non-stellar in origin since the absorption arises from the same hydrogen that causes the break.
%These features provide a critical window into the kinematic nature of the hydrogen present in the LRD system. 
The features are critical probes of the kinematic nature of LRDs which can be inferred from the velocity offset of the absorption. %Such absorption has been noted in $<1\%$ of Sloan Digital Sky Survey Quasars \citep{Wang2015} but is observed with increased frequency in LRDs. Selection effects are less likely to impact the SDSS sample (R = ???) but 
% if the absorption features are sufficiently narrow, they are likely lost in low-resolution NIRSpec data (prism, R = 30-300) and potentially lost in medium-resolution NIRSpec data ($R \sim 1000$). 

% The fraction of LRDs with Balmer absorption lines is an important constraint on both hydrodynamical models and radiative-transfer calculations currently being developed to model the LRDs. 
Many Balmer-line absorber LRDs have been identified in medium-resolution spectroscopy survey data \citep{deGraaff2025, Hviding2025, Mathee2024, Taylor2024, Kocevski2024}. Other surveys recover absorption in medium and high-resolution spectroscopy of single sources \citep{DEugenio2025TWICE, DEugenio2025, Kokorev2025,Lambrides2025, Naidu2025, Taylor2025, Torralba2025, Wang2025} while \citet{Matthee2026} and \citet{DEugenio2026} identified multiple Balmer-line absorbers for the first time in a high-resolution spectroscopy survey with the BlackTHUNDER Large Programme (Black holes in THe early Universe aNd their DensE surRoundings; program ID 5015; PIs: H. Übler, R. Maiolino). % Here, we expand the known Balmer-line absorbing LRD population through a new \textit{JWST} high-resolution spectroscopy survey to further probe this fraction of absorbing LRDs and explore  absorption as it may relate to other LRD properties.
Although current samples remain small and heterogeneous, the reported Balmer-line absorbers appear to lie preferentially among the reddest LRDs \citep{Barro2025, Matthee2026}. Here, we expand the known Balmer-line absorbing LRD population through a new high-resolution spectroscopy survey to assess whether this apparent color dependence holds and to explore how absorption relates to other LRD properties.

In this work, we use high-resolution ($R \sim 2700$) NIRSpec spectroscopy from the Observing Cosmic Evolution Among Nascent Systems (OCEANS) survey to identify and quantify this Balmer absorption and the presence of exponential wings in the emission lines of LRDs. In Section \ref{sec:Data}, we describe the OCEANS program, the low-resolution spectroscopy programs that targeted the OCEANS LRDs, which we utilize for continuum measurements, and the photometric coverage used to identify the LRDs. %In Section \ref{sec:selec_and_class}, we describe our photometric LRD color-color and compactness selection and line profile measurements. 
In Section \ref{sec:abs_properties}, we explore selection effects for Balmer-line absorption and explore trends with net-to-narrow emission, UV-to-optical color, and Balmer break strength. We present a case of velocity offset Balmer-line absorption and an accompanying cartoon model.
Throughout this work, we assume a flat $\Lambda$CDM cosmology consistent with measurements from Planck with $H_0 = 67.4$ $\mathrm{km} \mathrm{s}^{-1} \mathrm{Mpc}^{-1}$ and $\Omega_\mathrm{M}$ = 0.315 \citep{Planck2020}.
All logarithms are base ten. Redshift is denoted by z. Equivalent Widths (EWs) for absorption features are reported as a positive value from the rest-frame. All magnitudes (mag) are AB \citep{Osterbrock1989}.

%\citep{Pang2026}: cite for measuring broad + narrow lines

\section{Data}
\label{sec:Data}

% \subsection{NIRSpec Spectroscopy}
% \label{sec:spectroscopy}

We utilize NIRSpec spectroscopy from the high-resolution (G235H + G395H) OCEANS survey to probe the nature of Balmer-absorbing and non-absorbing LRDs. The CEERS photometry was used both to design the OCEANS MSA layout and select the LRDs in this work. We describe the data and selection methodology here.  We also utilize lower-resolution spectroscopy when available to measure the continuum and for comparison to the high-resolution spectral measurements.

\subsection{OCEANS High-Resolution Spectroscopy Survey}
\label{sec:spectroscopy_highres}

\noindent
\textbf{\underline{OCEANS}} We use high spectral resolution JWST/NIRSpec multi-shutter assembly (MSA) spectroscopy from the Cycle 4 Observing Cosmic Evolution Among Nascent Systems (OCEANS) survey [GO-8410, PI: R.C. Simons].

OCEANS consists of 6 NIRSpec MSA pointings in the Extended Groth Strip (EGS), targeting regions with existing NIRCam IR imaging from the CEERS survey \citep{Finkelstein2025} and optical-to-near-IR {\emph{HST}} imaging from the CANDELS survey \citep{Grogin11, Koekemoer11}.

NIRCam photometric data from the CEERS survey was used in the selection and classification of the LRDs and in targeting for the OCEANS survey. The photometric data used here is described in \citet{Bagley2023}. We use the CEERS photometric catalog version 0.51.2 and photometric redshifts presented in \citet{Finkelstein2024}. At the time of this survey, CEERS has photometric coverage in the rest-IR from the NIRCam filters F115W, F150W, F200W, F277W, F356W, F410M, and F444W in the Cycle 1 footprint. 

In parallel, OCEANS gathered F115W, F150W, F356W, and F410M NIRCam imaging in regions flanking the Extended Groth Strip (EGS). A photometric catalog and reduced images for the NIRCam component of OCEANS will be the subject of future work (Davis et al., in prep).

The spectroscopic targets for the MSA in OCEANS were selected from NIRCam imaging from the CEERS (ERS-1345; \citealt{Finkelstein2025}) survey and NIRSpec prism/medium-grating spectroscopy from the CEERS, CAPERS (GO-6368; PI: M. Dickinson),  THRILS (GO-5507; PIs: T. Hutchison and R. Larson; \citealt{Hutchison2025}), and RUBIES (GO-4233; PIs: A. de Graaff and G. Brammer; \citealt{deGraaff2025}) surveys. The primary targets were selected to enable spatially-resolved kinematic and outflow measurements at high redshift 
($2 \lesssim z \lesssim 7$; Simons et al. in prep). Of particular relevance to this work, the masks also include a high-priority sample of both confirmed and candidate LRDs and broad-line Active Galactic Nuclei (AGN), for which OCEANS provides the first high-resolution rest-optical spectroscopy. This enables the study of outflow kinematics from the LRD central engine presented here that are inaccessible in most systems at lower resolution. 
 
The LRD and AGN targets were compiled and prioritized by members of the OCEANS team on the basis of either (1) prior spectroscopic confirmation of broad-line emission \citep{Taylor2024} or (2) photometric identification as an LRD candidate with no prior spectroscopic observations \citep{Kocevski2024}. The selection also prioritized LRDs with previously identified absorption lines in CEERS. The majority of the LRD/AGN targets were observed with a single MSA pointing. Four of the targets discussed here were observed in multiple OCEANS pointings, detailed in the Appendix. For these sources we co-add the 1D extractions to decrease the signal to noise and improve our line fits. We exclude the OCEANS pointing 2 observation from co-adds because of a single unexpected closed shutter.                                                                                                           

%Sources were placed on 1-3 masks. To accomplish the goal of spatially resolving galaxy kinematics, targets were placed on either the leftmost or rightmost NIRSpec position so that subsequent observations would ????. Targets were selected for slits if they had spectroscopically confirmed AGN activity from \citep{Taylor2024}, if they had photometric properties that classified them as LRDs from \citet{Kocevski2024}, if they were photometrically selected as extreme emission line galaxies (EELGs) in \citet{Davis2023} with observed frame \OIII + \Hb\ EWs >$5000$ \AA, with priority given to EELGs at $z>7$, list other targets with sources...... also dusty dwarfs, some other galaxies, need full list from Raymond.
%Priority was given to.... describe how sources were targeted (info from Raymond).

%Give some info on internal reductions (raymond or madisyn?)

All targeted sources were observed using 2-shutter-long slitlets (total spatial extent $0\farcs92$), and the observations executed a 2-position nod pattern that placed the target in each shutter sequentially. Given that the high-redshift primary targets ($z>2$) are generally compact enough to fall within a single centered MSA shutter, the 2-shutter slitlets provide an ideal balance for estimating the local sky and maximizing the MSA multiplexing efficiency. Each OCEANS MSA pointing typically includes 70--90 science targets alongside 10 dedicated ``background" slitlets.

Each OCEANS MSA pointing was observed with both the G235H/F170LP and G395H/F290LP grating-filter combinations, providing continuous spectral coverage from $1.7$--$5.1~\mu$m at $R \sim 2700$. Science exposures consist of 10 integrations of 8 groups each for a total exposure time of 14,734 s ($\sim$ 4.1 h) per grating per pointing. Target acquisition was performed via the MSA target acquisition (\texttt{MSATA}) method using the F140X filter with 6--8 suitable reference stars distributed across the MSA field of view. No additional spatial dithers were employed, with the 2-position nod serving as the only positional stepping between exposures. Of relevance to this paper, the P2 pointing of OCEANS-35829 (targeted by 3 total pointings) experienced a single unexpected closed shutter. As a result, the 1D spectrum for this source is extracted from a single nod position. The six pointings were observed across a range of aperture position angles ($5\arcdeg \lesssim \mathrm{APA} \lesssim 315\arcdeg$).

We reduce the NIRSpec MOS data using the {\emph{JWST}} calibration pipeline (v1.20.2, CRDS context jwst\_1464.pmap; \citealt{bushouse_2025_16280965}), adapting the NIRSpec MOS pipeline notebook \citep{jwst_pipeline_notebooks} with custom modifications for cosmic ray rejection and bad pixel self-calibration. In Stage 1 (\texttt{calwebb\_detector1}), we adopt the enhanced snowball correction parameters used by the CEERS survey, with an expansion factor of 3 and a minimum flagged area of 15 pixels. In Stage 2 (\texttt{calwebb\_spec2}), we enable bad pixel self-calibration (\texttt{badpix\_selfcal}; $f_\mathrm{flag} = 0.005$) and apply standard nodded background subtraction. Final combined spectra were produced in Stage 3 (\texttt{calwebb\_spec3}) by co-adding the two nod positions for each source. 1D spectra ({\tt{x1d}}) were extracted using a 5-pixel wide box-car aperture centered on the source. For each source, we measure the empirical pixel-level noise as 1.486$\times$ NMAD of the continuum subtracted 1D spectrum and compare it to the median of the pipeline-reported uncertainties. We find that the pipeline systematically underestimates the noise by $\sim35-65\%$ across the sample (mean 45$\%$). For each source, we rescale the 1D uncertainty array by its empirical correction factor.

The OCEANS spectroscopy achieves an empirical $5\sigma$ line flux sensitivity of $\sim$2 -- 4~$\times 10^{-19}$~erg~s$^{-1}$~cm$^{-2}$ across the G235H bandpass and $\sim$1 -- 3~$\times 10^{-19}$~erg~s$^{-1}$~cm$^{-2}$ across G395H, assuming a line width of $\sigma = 50$~km~s$^{-1}$, with a sensitivity minimum near $\lambda_{\rm obs} \sim 4.0~\mu$m in G395H. The empirical sensitivity is derived from a median error spectrum from each pointing, scaled by the empirical underestimation of the pipeline errors.

\subsection{OCEANS LRD Photometric Sample Selection}
\label{sec:select}

\begin{figure}[h]
    
    \centering
    \includegraphics[width=1\linewidth]{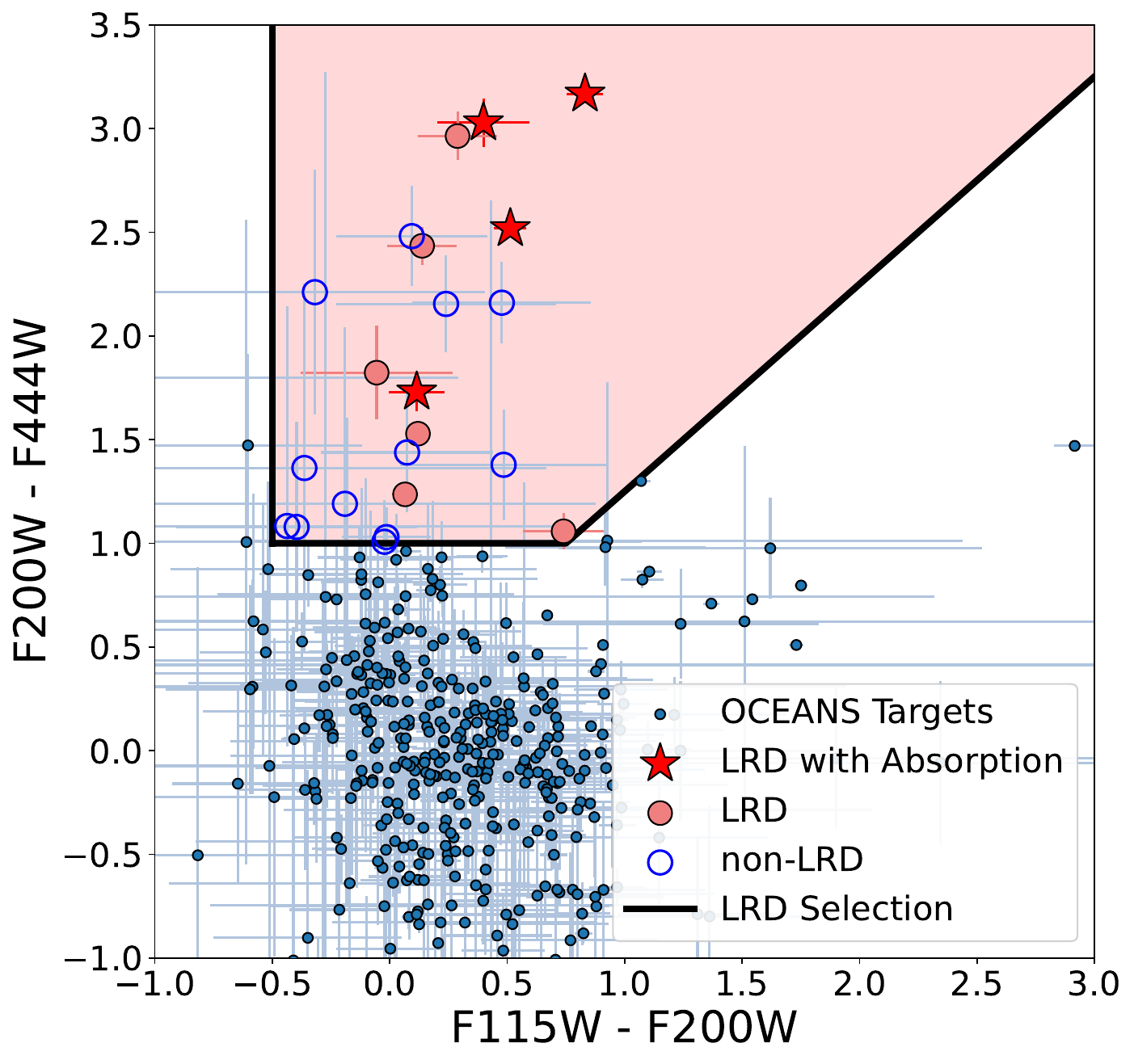}
    \caption{ The OCEANS LRD sample in color-color space. All OCEANS targets are shown in solid light blue, OCEANS LRDs with broad \Ha\ emission are shown as coral circles, and OCEANS LRDs with broad Balmer emission and absorption features are indicated with red stars. Blue circles indicate OCEANS sources with LRD colors that did not pass the compactness selection in Figure \ref{fig:compact} or the $\rm{F444W <27 }$ (mag) constraint. %Contours represent 1 $\sigma$,  2 $\sigma$ and 3 $\sigma$ population contours for galaxies in CEERS at $2<z<9$. 
    The black line and red shaded region shows the color-color selection from \citet{Barro2026} for LRDs.  
    }
    \label{fig:colorcolor}
    
\end{figure}

\begin{figure}[h]
   
    \centering
    \includegraphics[width=1\linewidth]{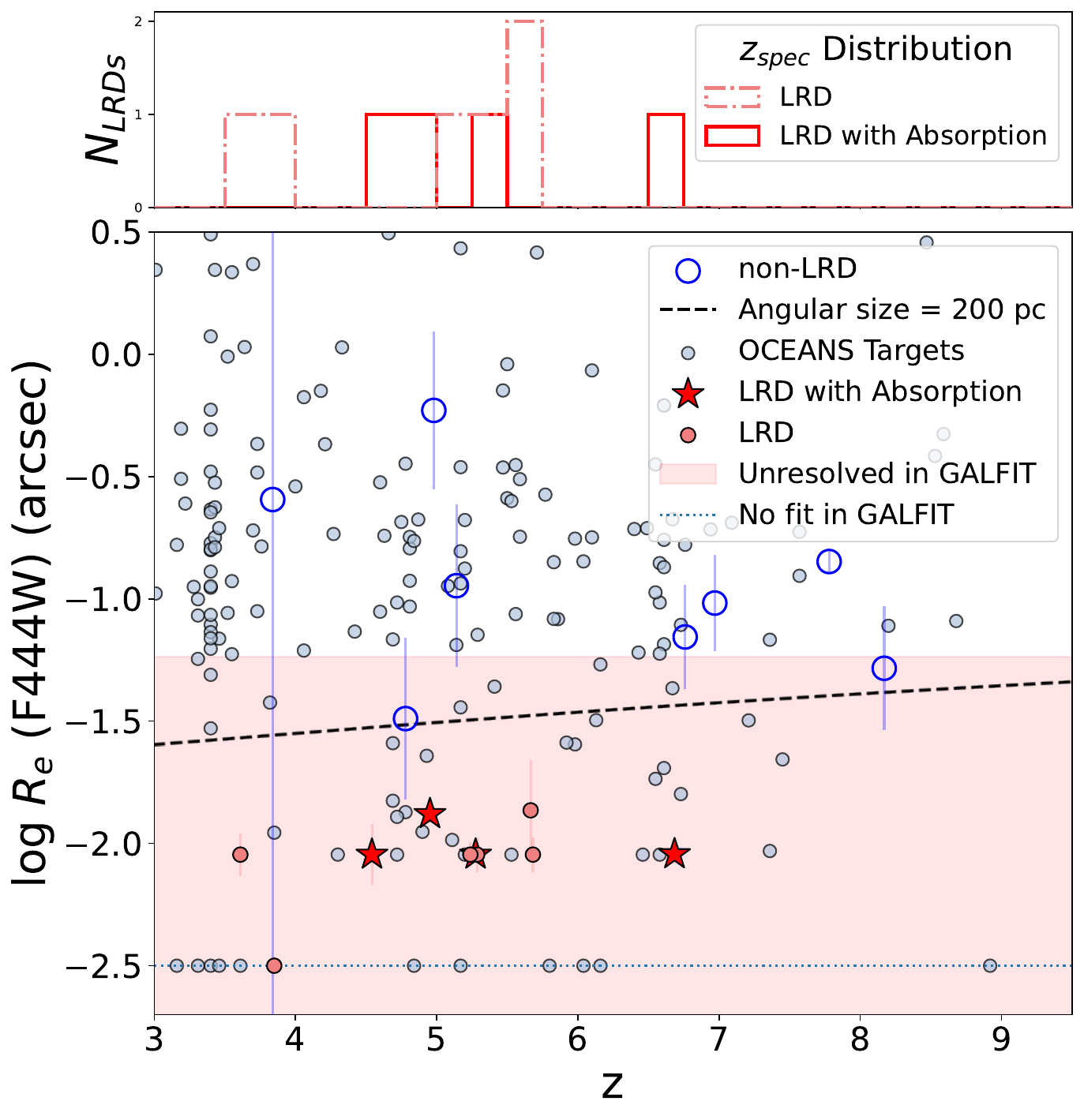}
    \caption{\texttt{GALFIT} measurements from \citet{McGrath2026} for CEERS and the LRD population. LRDs are plotted as coral circles while OCEANS LRDs with Balmer absorption are plotted as red stars. We use these half-light radii ($R_e$) to apply a compactness cut associated with an angular size of $<200$ pc and ensure our color selected LRDs are compact. We mark sizes smaller than the HWHM of the F444W PSF by the red shaded region and call sources that fall below this limit ``unresolved.'' 
    This approximately correlates with our angular size cut, consistent with the LRD representing unresolved point sources. Other oceans targets that are not color selected are plotted as solid blue circles. %but do not apply a quality flag constraint to the LRDs.
    }
    \label{fig:compact}
\end{figure}

We select our sample of LRDs by defining the sources in the photometry to retain consistency with the original selection of LRDs and ensure no targets from the OCEANS survey with LRD properties are missed. % that identified them as outliers in the photometry. 
Our sample is selected with the color cuts described in \citet{Barro2026} and further required LRDs to have compact half-light radii ($R_e$).

We first apply the selection in \citet{Barro2026} that defined the LRD color-color selection as  $\rm{F200W} - \rm{F444W > 1}$, $(\rm{F200W} - \rm{F444W}) > (\rm{F115W} - \rm{F200W})+0.25$, $\rm{F115W} - \rm{F200W > -0.5}$, and limit of $\rm{F444W <27 }$ mag. We show our LRD sample selection with all other OCEANS targets in Figure \ref{fig:colorcolor}. Sources that met LRD color criteria for compactness were checked against the F444W $R_e$, requiring that sources have angular sizes $<200$pc. This kept our definition of LRDs consistent with established rest-UV and rest-optical colors combined with optical compactness \citep{Kocevski2024}. 

We use \texttt{GALFIT} measurements from \citet{McGrath2026} of $R_e$ for our compactness cut in Figure \ref{fig:compact}. Our expectation is that LRDs are highly compact in rest-optical light and may be unresolved in the photometry.
We indicate an unresolved $R_e$ with the red shaded region in Figure \ref{fig:compact}, the half-width-half-max (HWHM) of the point spread function (PSF) for the F444W photometric filter. We also indicate an angular size corresponding to 200~pc with a black dashed line. This corresponds roughly with the HWHM of an unresolved source in F444W, consistent with this compactness threshold translating to unresolved point sources in F444W.
We allow any flag in \texttt{GALFIT} to return a result so that we can recover sources that are unresolved in rest-optical light but we visually inspect all postage stamps to confirm compactness.

The color-color selection identifies 22 candidate LRDs. When the magnitude constraint ($\rm{F444W <27 }$ mag) is applied, 9 LRDs are removed. The compactness cut removes an additional LRD, OCEANS 33961, that passes the color and mag cuts. All removed sources are marked with open blue circles in Figures \ref{fig:colorcolor} and Figure \ref{fig:compact}. All potential targets were investigated for broad Balmer lines present in all true LRDs \citep{Hviding2025} and none were found.
We exclude OCEANS 33082 which has evidence for broad \Ha\ but no reference forbidden lines or low-resolution spectroscopy. The source was selected from the photometry in \citet{Kocevski2024} and is likely a genuine LRD, but we cannot verify the broad-line AGN nature without reference narrow lines. We also remove OCEANS 110839 because of a slit failure.

All remaining LRDs in the sample are also photometrically selected as LRDs by \citet{Kokorev2024} and/or \citet{Kocevski2024} except for OCEANS 169045 and 1794. OCEANS 169045 was identified as an early low-luminosity AGN in \citet{Kocevski2023} and OCEANS 1794 has a similar line profile shape. These sources pass all previously mentioned cuts and therefore pass our defined LRD criteria and we include them in the sample.

Our color selection selects 22 LRDs, 9 of which are lost to the magnitude cut. An additional source is lost to the compactness cut and one source is removed for insufficient line coverage while another is removed because the observation failed. This leaves 10 LRDs 4 of which are best fit with absorption components. These numbers are given in Table \ref{tab:lrd_selection}.

\begin{table}[h]
\centering
\caption{LRD sample selection by cut}
\label{tab:lrd_selection}
\begin{tabular}{lc}
\hline
\hline
Selection Step & N LRDs \\
\hline
Color-color selection & 22 \\
$\rm{F444W <27 }$ mag) & $-9$ \\
 ($R_e < 200 \,\rm{pc}$) & $-1$ \\
Insufficient line coverage & $-1$ \\
Failed slit & $-1$ \\
\hline
Final LRD  sample & 10 \\
LRDs with Absorption & 4 \\
\hline
\hline
\end{tabular}
\end{table}

\subsection{Archival Lower-Resolution Spectroscopy}
\label{sec:spectroscopy_lowres}

Low-resolution ($R\sim100$) prism spectroscopy from NIRSpec programs targeting CEERS are utilized in this work to measure the rest-optical and rest-UV continuum of the LRDs. We describe the overlapping surveys in brief here.

\noindent
\textbf{\underline{CAPERS}} The CANDELS-Area Prism Epoch of Reionization Survey (CAPERS) [program ID 6838, PI: Mark Dickinson] target selection and reduction are described in \citet{Kokorev2025, Taylor2025}. 
The CAPERS program included 7 pointings of NIRSpec spectroscopy in EGS/CEERS with $1-3$ micro-shutter array (MSA) configurations and exposure times of $1.58-4.74$ hours. This prism spectroscopy has a low resolution of $R \sim 100$.

\noindent
\textbf{\underline{RUBIES}} We include NIRSpec data from the Red Unknowns: Bright Infrared Extragalactic Survey (RUBIES) [program ID 4233, PIs: A. de Graaff \& G. Brammer] \citep{deGraaff2025}, utilizing reductions from \citet{Taylor2024}. 
The RUBIES program included 6 NIRSpec pointings in EGS/CEERS using both low-resolution prism spectroscopy (R $\sim $100) and medium-resolution G395M spectroscopy (R $\sim 1000$) with 0.8 hour exposures in both modes. We use the medium-resolution spectra for resolution comparisons and the low-resolution spectra for continuum measurements.

We measure continuum properties for the LRDs from available low-resolution spectroscopy, always choosing the observation with the longest exposure time. We measure Balmer breaks as $f_{\nu, 4100}/ f_{\nu, 3600}$ and the UV-to-optical color as  $f_{\nu, 5500}/ f_{\nu, 3600}$. All fluxes for the colors are median flux densities taken from a rest-frame 300 \AA\ window centered on the target wavelength.

\section{LRD Profile Fitting}
\label{sec:selec_and_class}

\subsection{Balmer Line Profile Fitting}
\label{sec:wings}

We begin our line-fitting procedure by calculating a redshift from the peak of the \OIII$\lambda$5008 line to define the systemic redshift. For the Balmer emission lines, we fit with a narrow Gaussian whose width and line center we allow to vary, a narrow Gaussian whose parameters we tie to \OIII$\lambda$5008, an exponential  convolved with a Gaussian component (or optionally, a purely Gaussian component), and an absorption feature. The absorption features are fit by a Gaussian profile whose amplitude, width, and line center are allowed to vary. We set a minimum width of $40 \, \rm{km} \, \rm{s}^{-1}$ for the absorption feature. Our \Ha\ fits include potential contribution from the \NII$\lambda 6548,6583$ doublet with a fixed flux ratio of 3 \citep{Osterbrock1989}.

\begin{figure*}
    \centering
    \includegraphics[width=1\linewidth]{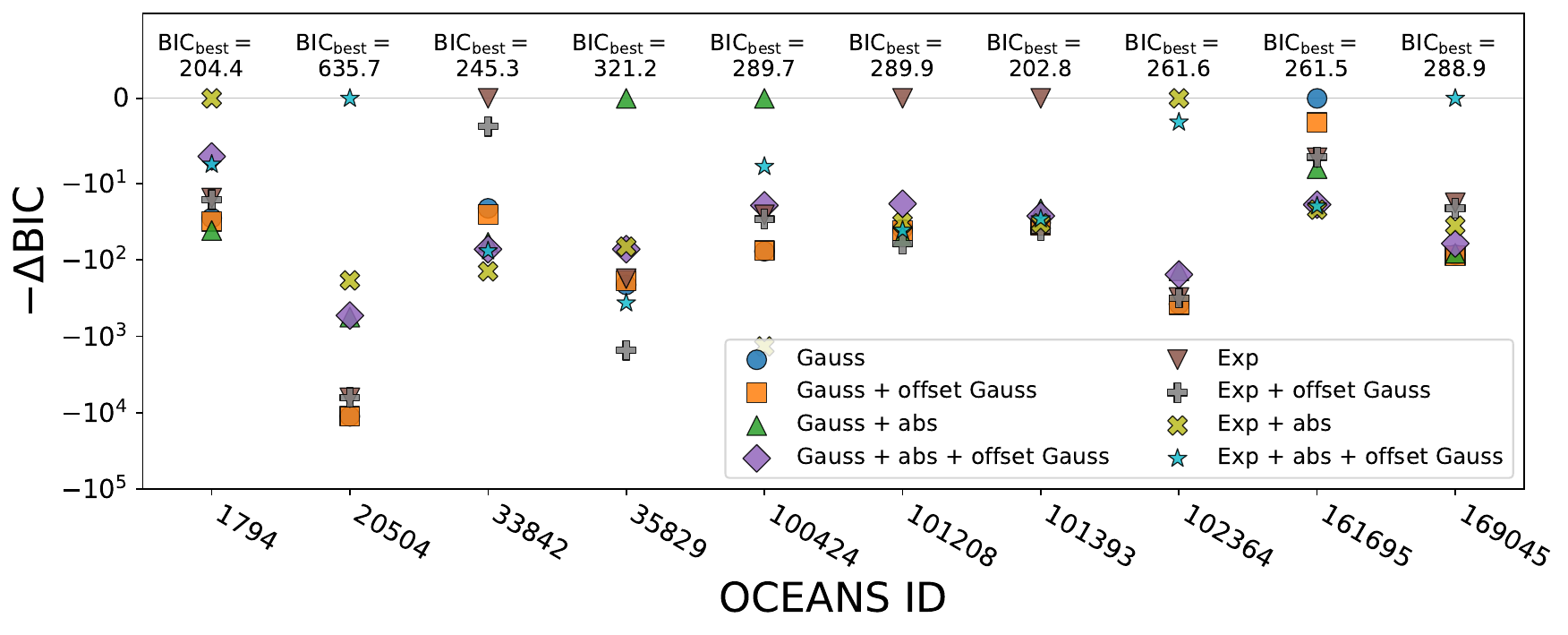}
    \caption{\Ha\ $\Delta\rm{BIC}$ by source for the OCEANS LRD set. Each possible combination of fit parameters is listed in the legend and assigned a unique marker. The $\Delta\rm{BIC}$ is plotted relative to the minimized $\Delta\rm{BIC}$ which best fit each source with the best-fit $\Delta\rm{BIC}$ plotted at 0. Gauss is short for Gaussian only (no exponential line wings), Exp indicates an exponential convolved with a Gaussian was the best fit. Offset Gauss indicates sources where the broad centroid is offset from 0 velocity. Abs indicates a fit favors the addition of an absorption component. The absorption fits favored for OCEANS 169045 and 1794 are $\Delta\rm{BIC}$ artifacts, since the sources does not have any signs of absorption in their line profiles. The next best fit without absorption (exponential only) fit is assumed in this work. OCEANS 101393 has potential near-zero-velocity weak absorption (see the insets for all sources in Figure \ref{fig:absorbers} and \ref{fig:noabs}) but the $\Delta\rm{BIC}$ does not favor an absorption feature and the visual dips in flux span a few pixels. We exclude it from the absorbers. }
    \label{fig:dbic}
\end{figure*}
% We show the fits that minimize the $\Delta \rm{BIC}$, parameter which we define as….., in the body of this text and give fit details in the Appendix. We find that all but one of the Balmer-absorbing LRDs are best fit with exponential Gaussians.

Our combined narrow and broad line profiles, $f(\lambda)_{dual}$, are described by Equation \ref{eqn:comb} where $f_{c}$ is the local continuum, $f_{nar}$ is the narrow-only line component, and $f_{broad}$ is the broad Gaussian amplitude. $\sigma_{nar}$ and $\sigma_{broad}$ are the narrow and broad line widths.

\begin{equation}
\begin{split}
    f(\lambda)_{dual} = f_{c} + f_{nar} \exp{\bigg(-\frac{1}{2}\frac{(\lambda - \lambda_{0,narrow})^2}{\sigma^{2}_{nar}}}\bigg) \\
    + f_{broad} \exp{\bigg(-\frac{1}{2}\frac{(\lambda - \lambda_{0,broad})^2}{\sigma^{2}_{broad}}}\bigg)
    \label{eqn:comb}
\end{split}
\end{equation}

In the case that an exponential line shape is prefered, the second term in \ref{eqn:comb} takes the form of an exponential convolved Gaussian described in Equation \ref{eqn:expformula} where $\gamma$ is a scale parameter defined within the double exponential function $f(\lambda)=(1/2\gamma) * e^{-x|(\lambda - \lambda_0 ) / \gamma|}$ which is then convolved with a broad Gaussian feature. Our definitions are consistent with those from other exponential wing fitting papers \citep{Matthee2026, Rusakov2026, Kokorev2025}.

\begin{equation}
% \begin{centering}
\begin{split}
f(\lambda)_{broad} = 
\sqrt{\frac{\pi}{2}} \frac{f_{broad}}{2} 
\frac{\sigma_{broad}}{\gamma_{wings}}
\Bigg[
\exp\left(\frac{\sigma_{broad}^2}{2\gamma_{wings}^2}
- \frac{(\lambda - \lambda_{0})}{\gamma_{wings}}\right) \\ \times
~\text{erfc}\left(
\frac{\sigma_{broad}^2 - (\lambda - \lambda_{0})\gamma_{wings}}
{\sqrt{2}\sigma_{broad}\gamma_{wings}}
\right) 
\quad \\ + \exp\left(\frac{\sigma_{broad}^2}{2\gamma_{wings}^2}
+ \frac{(\lambda - \lambda_{0})}{\gamma_{wings}}\right) \\ \times
~\text{erfc}\left(
\frac{\sigma_{broad}^2 + (\lambda - \lambda_{0})\gamma_{wings}}
{\sqrt{2}\sigma_{broad}\gamma_{wings}}
\right)
\Bigg]
\end{split}
% \end{centering}
\label{eqn:expformula}
\end{equation}

\begin{figure*}[h]
\begin{center}
    \includegraphics[width=1\linewidth]{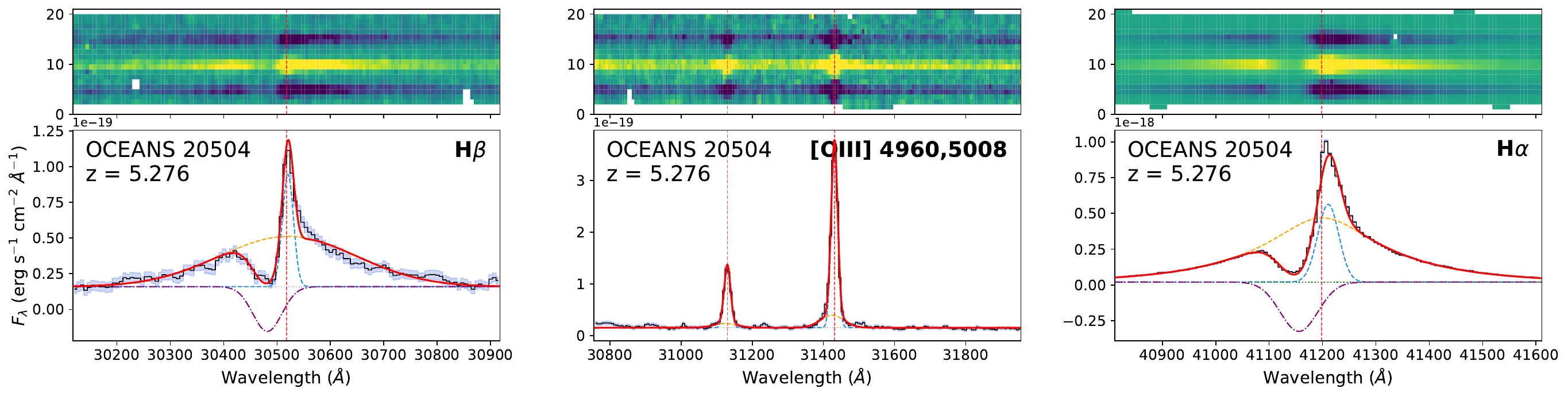}
    \includegraphics[width=1\linewidth]{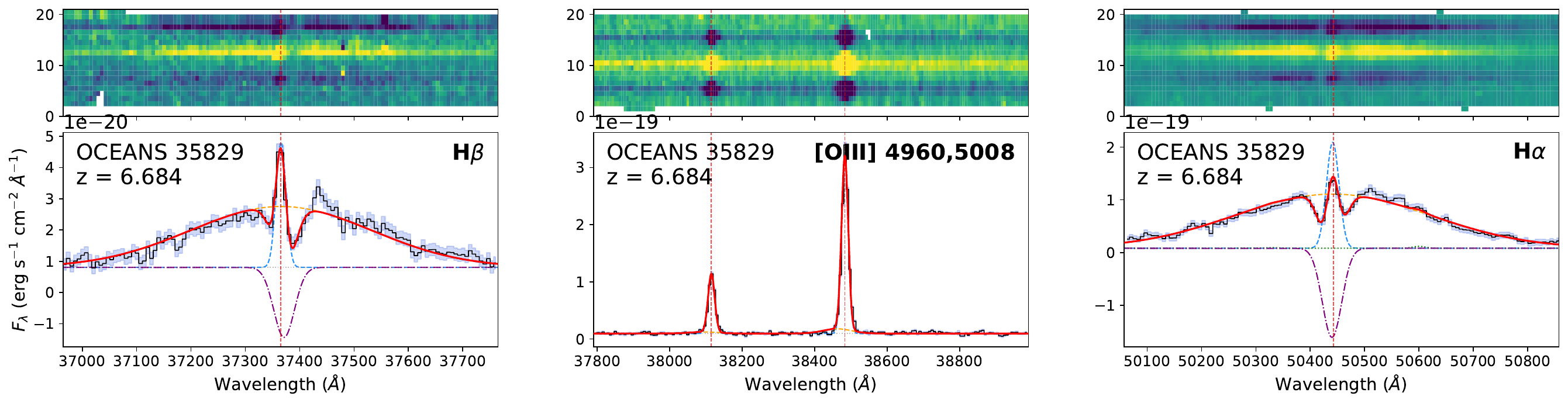}
    \includegraphics[width=1\linewidth]{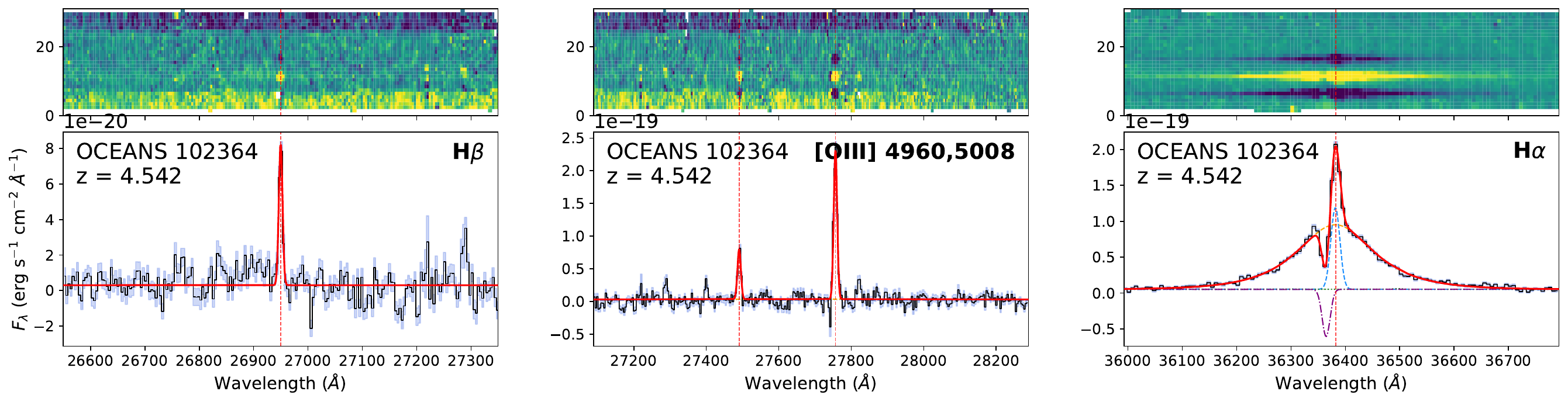}

    \includegraphics[width=1\linewidth]{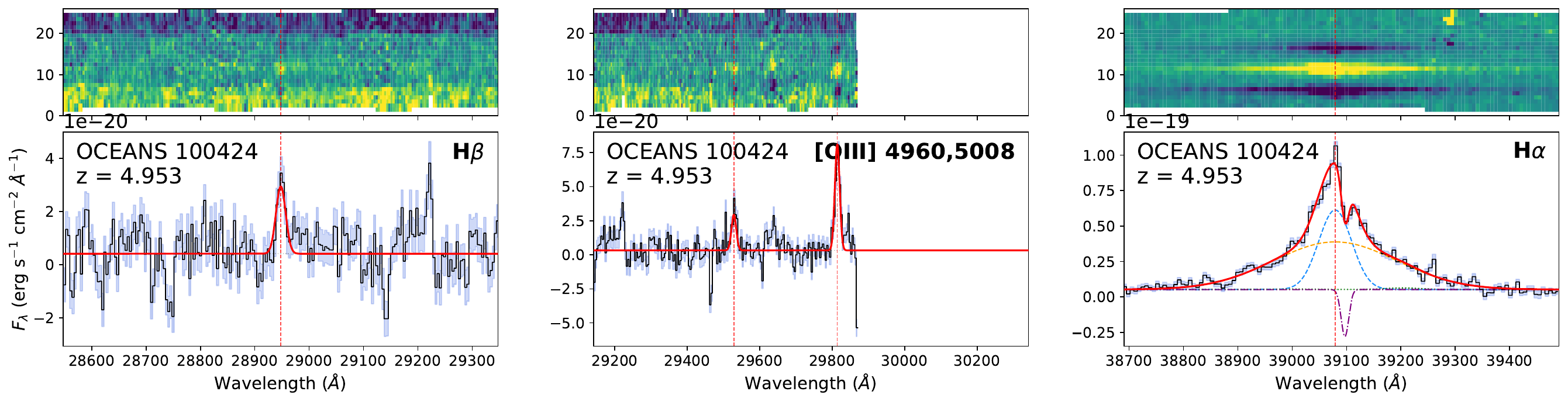}

    \includegraphics[width=1\linewidth]{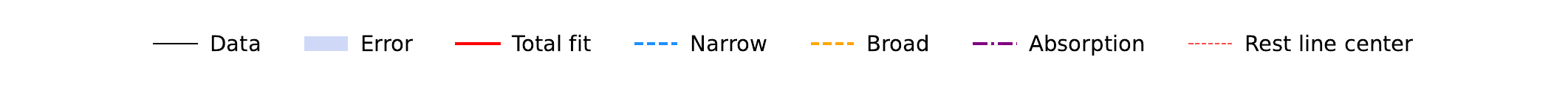}

    % \includegraphics[width=0.31\linewidth]{s000020504_G395H_Halpha.pdf}
    % \includegraphics[width=0.31\linewidth]{s000020504_G395H_Hbeta.pdf}
    % % \includegraphics[width=0.24\linewidth]{s000020504_G235H_Hgamma_and_OIII4364.pdf}
    % \includegraphics[width=0.31\linewidth]{s000020504_G395H_OIII_4960-5008.pdf}
    
     %.......................................
    %  \includegraphics[width=0.31\linewidth]{s000035829_G395H_Halpha.pdf}
    %  %\includegraphics[width=0.31\linewidth]{hb35829.png}
    %  \includegraphics[width=0.31\linewidth]{s000035829_G395H_Hbeta.pdf}
    % % \includegraphics[width=0.31\linewidth]{s000035829_G395H_Hgamma_and_OIII4364.pdf}
    % \includegraphics[width=0.31\linewidth]{s000035829_G395H_OIII_4960-5008.pdf}
    
    % \includegraphics[width=0.31\linewidth]{s000102364_G395H_Halpha.pdf}
    % \includegraphics[width=0.31\linewidth]{s000102364_G235H_Hbeta.pdf}
    % % \includegraphics[width=0.24\linewidth]{s000102364_G235H_Hgamma_and_OIII4364.pdf}
    % \includegraphics[width=0.31\linewidth]{s000102364_G235H_OIII_4960-5008.pdf}
    
    % \includegraphics[width=0.31\linewidth]{s000101393_G395H_Halpha.pdf}
    % \includegraphics[width=0.31\linewidth]{s000101393_G235H_Hbeta.pdf}
    % % \includegraphics[width=0.24\linewidth]{s000101393_G235H_Hgamma_and_OIII4364.pdf}
    % \includegraphics[width=0.31\linewidth]{s000101393_G235H_OIII_4960-5008.pdf}
    
    % \includegraphics[width=0.31\linewidth]{s000100424_G395H_Halpha.pdf}
    % \includegraphics[width=0.31\linewidth]{s000100424_G235H_Hbeta.pdf}
    % % \includegraphics[width=0.24\linewidth]{s000100424_G235H_Hgamma_and_OIII4364.pdf}
    % \includegraphics[width=0.31\linewidth]{s000100424_G395H_OIII_4960-5008.pdf}

    \caption{LRDs with \Ha\ absorption features from OCEANS. Each row in the figure represents a single source in OCEANS. First column insets are \Hb, second are \OIII$\lambda$4960,5008, and third are \Ha. The OCEANS spectrum is plotted in black and error is plotted as the light blue shaded region. Each inset has the total fit that gives the preferred $\Delta \rm{BIC}$ described in Section \ref{sec:Data}. Narrow \Ha\ is plotted with a light blue dotted line, broad Gaussian or exponential profiles are plotted as orange dashed lines, \NII\ contribution is plotted as a green dashed line, and absorption is plotted as a purple dot-dashed line. Insets for non-absorbing LRDs can be found in Figure \ref{fig:noabs}.}
    \label{fig:absorbers}
    
\end{center}
\end{figure*}

\begin{figure*}
    \begin{center}

    \includegraphics[width=.78\linewidth]{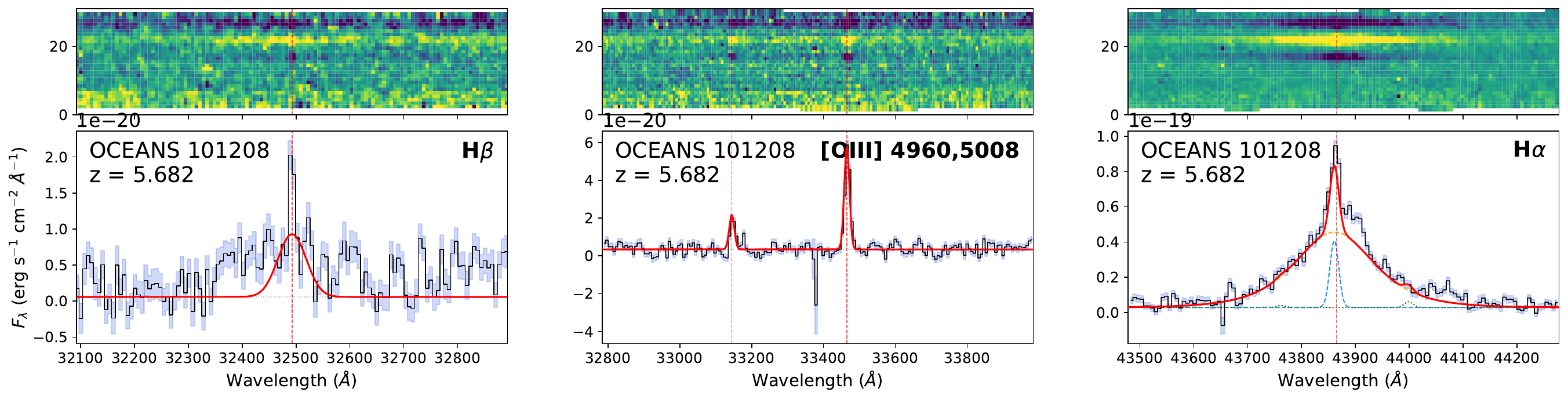}
    \includegraphics[width=.78\linewidth]{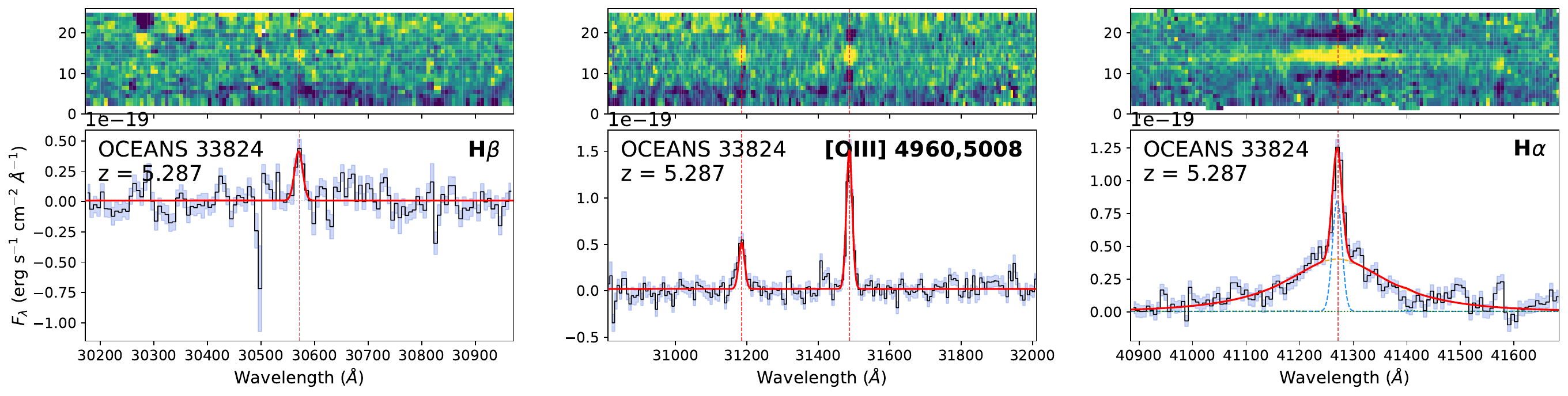}
    \includegraphics[width=.78\linewidth]{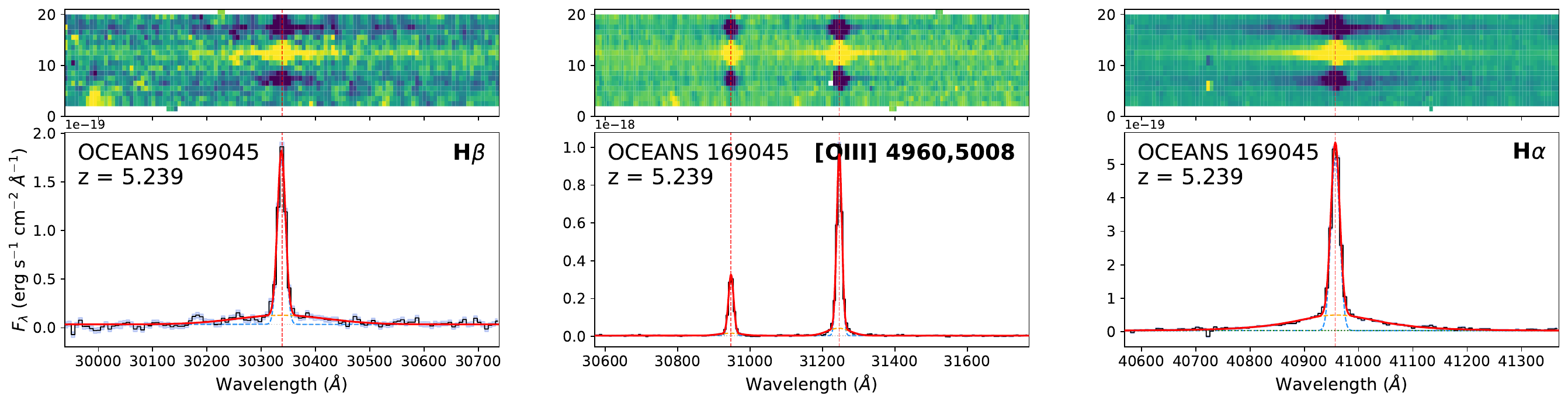}
    \includegraphics[width=.78\linewidth]{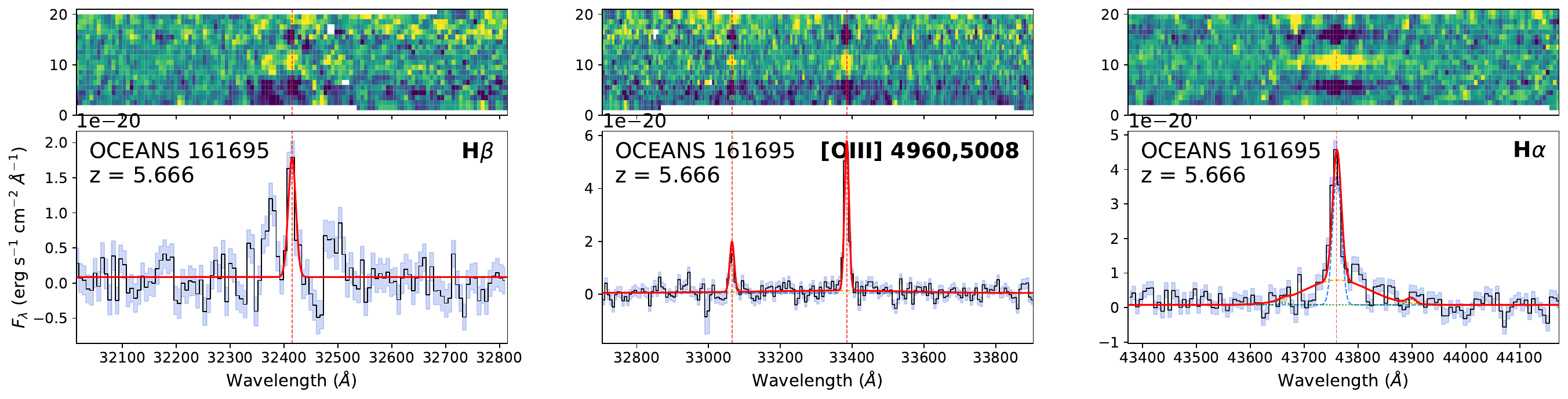}
    \includegraphics[width=.78\linewidth]{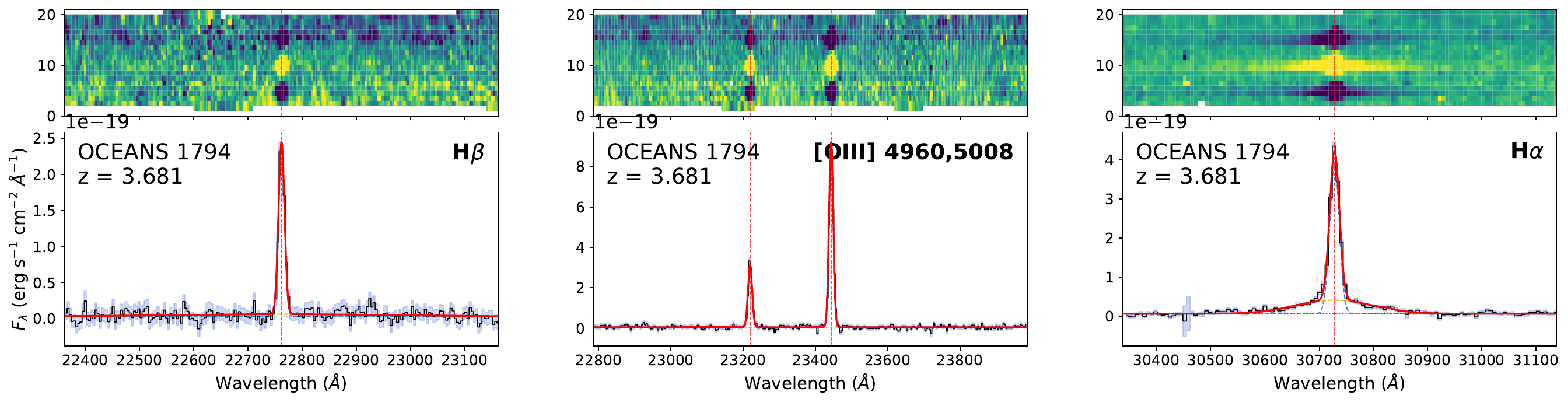}
    \includegraphics[width=.78\linewidth]{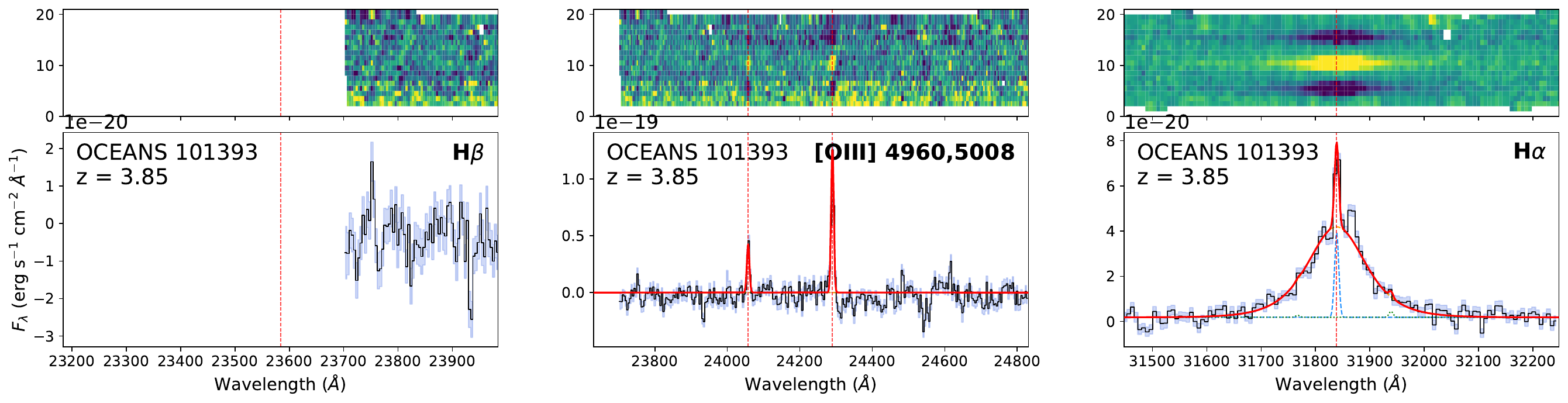}
    \includegraphics[width=.78\linewidth]{inset_legend.pdf}

    \caption{LRDs with broad \Ha\ emission and no absorption from OCEANS. Each row in the figure represents a single source in OCEANS. First column insets are \Hb, second are \OIII$\lambda$4960,5008, and third are \Ha.. The OCEANS spectrum is plotted in black and error is plotted as the light blue shaded region. Each inset has the total fit that gives the preferred  $\Delta \rm{BIC}$ described in Section \ref{sec:Data}. Narrow \Ha\ is plotted with a light blue dotted line, broad Gaussian or exponential profiles are plotted as orange dashed lines, \NII\ contribution is plotted as a green dashed line, and absorption is plotted as a purple dot-dashed line. Redshift and source IDs are listed in the legends.}
    \label{fig:noabs}
    \end{center}
\end{figure*}

We fit the \OIII$\lambda4960, 5008$ lines by allowing two narrow Gaussians of fixed flux ratio 2.98 (from \citet{Osterbrock1989}) and two broad Gaussian components which we allow to vary in line center. Figures \ref{fig:absorbers} and \ref{fig:noabs} show that, while three LRDs (two of which are absorbers) have evidence for weak broad \OIII, the lines are much weaker and narrower than the broad \Ha\ components. %This may point to galactic-scale outflows in some LRDs, but 
None of the LRDs that have broad \OIII\ are comparable in strength to the broad \Ha. Our results are consistent with previous studies of BLAGN observed in the EGS field \citep{Kocevski2023, Taylor2024, Brooks2025}.

% We utilize the Bayesian information criterion (BIC) to compare model goodness-of-fit, where BIC is defined as $BIC = \chi^2 + k \ln(N)$,  $k$ is the number of model free parameters, and $N$ is the number of data points included in the fits. 
% We evaluate a range of model parameters of all fit components and report the fit that minimizes the difference between the BIC fit to the combined components and to the model, the $\Delta \rm{BIC}$, and require a  $\Delta \rm{BIC}>6$ to accept one model over another. The narrow-line-only profile is our fiducial. We add  a broad \Ha\ Gaussian component, the \NII\ doublet, a broad \Ha\ exponential component, and an \Ha\ absorption line if the model  $\Delta \rm{BIC}$ justifies the addition

To determine the best fit \Ha\ model, we utilize the Bayesian information criterion (BIC) to compare model goodness-of-fit and determine if extra fit parameters (absorption, free broad line center, etc) are needed to best describe the broad \Ha line profile. Here, the BIC is defined as $BIC = \chi^2 + k \ln(N)$, where $k$ is the number of model free parameters, and $N$ is the number of data points included in the fits. We start our iterative fitting process with the fiducial narrow-line-only \Ha\ $+$ \NII$\lambda 6548,6583$ profile. We then add additional fit components as determined by the $\Delta \rm{BIC}$. We require a $\Delta \rm{BIC}>6$ to accept the added fit component and then update our ``fiducial" model. In total, each spectra is fit with 8 different \Ha\ line profile models, shown in Figure \ref{fig:dbic}. In total, an LRD could be fit with up to 12 free parameters to describe its broad \Ha\ profile; this 12-parameter model would include a narrow Gaussian, a broad Gaussian convolved with an exponential profile, and an absorption component. The broad and narrow component line centers are allowed to vary. 

All LRDs have  $\Delta \rm{BIC}$ evidence for broad \Ha\ emission. A complete list of the best models for each LRD and the  $\Delta \rm{BIC}$ values with respect to the best fit model for \Ha\ are given in Figure \ref{fig:dbic}. The line profiles for OCEANS sources with fits that have significant absorption components are shown in Figure \ref{fig:absorbers} and we show the remaining OCEANS LRD sample in Figure \ref{fig:noabs}. Our final sample includes 4 absorbers and 6 non-absorbers. 

We fit \Hb\ profiles with a narrow-only fiduciary model and include an optional broad \Hb\ component when the  $\Delta \rm{BIC}$ prefers the addition over the single Gaussian-only model. We include an absorption component in the \Hb\ fit, with the same parameters allowed for the \Ha\ profile, in the case that an absorption feature is visibly apparent. We require a $3 \sigma$ detection of a braod \Hb\ component.

% the narrow-line profile is the fiducial, and then you add a broad \Ha\ Gaussian component, the \NII\ doublet, a broad \Ha\ exponential component, and an \Ha\ absorption line if justified by dBIC>6. All LRDs have dBIC evidence for a broad Ha Gaussian component, and then list the numbers of LRDs that have evidence for the other components. The 

% Our final classification includes 5 absorbing LRDs and 7 LRDs without detected absorption features. The final sample has an absorption recovery rate of $42\%$. Of these, OCEANS 20504 was identified as an absorber at medium resolution as RUBIES 42046 in \citet{Taylor2024} and OCEANS 35829 was identified as an absorber also at medium-resolution as the ``GlimmIR'' or ``Irony'' source in \citet{Lambrides2025} and \citet{DEugenio2025}. 
% The remaining absorbers, OCEANS 102364, 101393, and 100424 are reported here for the first time. All were spectroscopically observed in the RUBIES and/or CAPERS surveys, but their absorption features were undetected at lower resolution. Other OCEANS LRDs without detected absorption have coverage from other programs (see the full list in the Appendix). OCEANS 161695 and 61557 are new broad-line AGN identified as LRDs from the photometry in \citet{Kocevski2024} and spectroscopically confirmed here for the first time. 

\subsection{Previous Observations of OCEANS LRDs in the Literature}

A few of the LRDs in our sample have been previously noted in the literature. OCEANS 169045 was originally identified as the low-luminosity AGN CEERS 2782 in \citet{Kocevski2023}. We confirm the source's broad \Ha\ emission here and show that, even in this LRD whose broad \Ha\ component is sub-dominant, the source is best fit with exponential wings. Additionally, we recover broad \OIII$\lambda5008,4960$ in this source indicating the host galaxy may be experiencing an outflow that was missed by previous lower resolution observations of the source in CAPERS, RUBIES, and CEERS programs. The broad \OIII\ in OCEANS 169045 is broad but not comparable in intensity to the broad \Ha\ feature, which is consistent with a galactic outflow broadening the line.
We do not recover Balmer absorption in the source. %and find a moderate Balmer break which points to star formation, rather than the host LRD, driving the outflow.

OCEANS 35829 was observed by The High-(Redshift+Ionization) Line Search (THRILS) [program ID 5507, PIs: T. Hutchison \& R. Larson] \citep{Hutchison2025} as THRILS 46403 and is a noted Balmer absorber discussed in \citet{Lambrides2025} and \citet{DEugenio2025}, which identified the absorption at medium spectral resolution. We recover Balmer absorption here from two OCEANS pointings of the source, two of which capture \Hb\ and one of which captures \Ha.

OCEANS 20504 was observed in the RUBIES program as RUBIES 42046 and was identified as a Balmer-absorbing LRD at medium-resolution \citep{Kocevski2024, Taylor2024}. We recover broad \OIII\ emission in this source that is much weaker and narrower than the broad \Ha\ component, pointing to a galactic outflow also present in the system.

\section{Absorber Properties}
\label{sec:abs_properties}

We explore the properties of Balmer-line absorption as tied to spectral resolution and the properties of the absorbers themselves.

\subsection{Selection Effects}

We contextualize Balmer-line absorbing LRDs by interpreting their detection through the lens of their spectroscopic resolution. We find a higher incidence of absorption features at higher resolution, note high-resolution data successfully recovers absorption lines in lower-luminosity LRDs, and report two low luminosity absorbing LRDs.

\subsubsection{Spectral Resolution and Absorption-Line Detection}
% degraff: balmer line profiles with strong ab near line center:do we not pick these up because of observational bias? Should cite her.

% I looked at other papers with LRDs that have absorption:
% J Greene: UCOVER \citep{Grazian2024} 0/17 absorbers at prism resolution 
% cliffs: taylor, naidu, prism resolution and get absorption
% mathee 2024 \citep{Mathee2024} found 2/20 or 10 prcnt
%\citep{Hviding2025} has absorption in figures but doesn't explicetly identify or give a count

% \citep{Taylor2024} found abs in 4/21 LRDs (we exclude non-LRD AGN) or 19 prcnt

% kocevski24 found 3/17

% \citep{Matthee2026} found 60 prcnt 4/7

\begin{figure}
    \centering
    \includegraphics[width=1\linewidth]{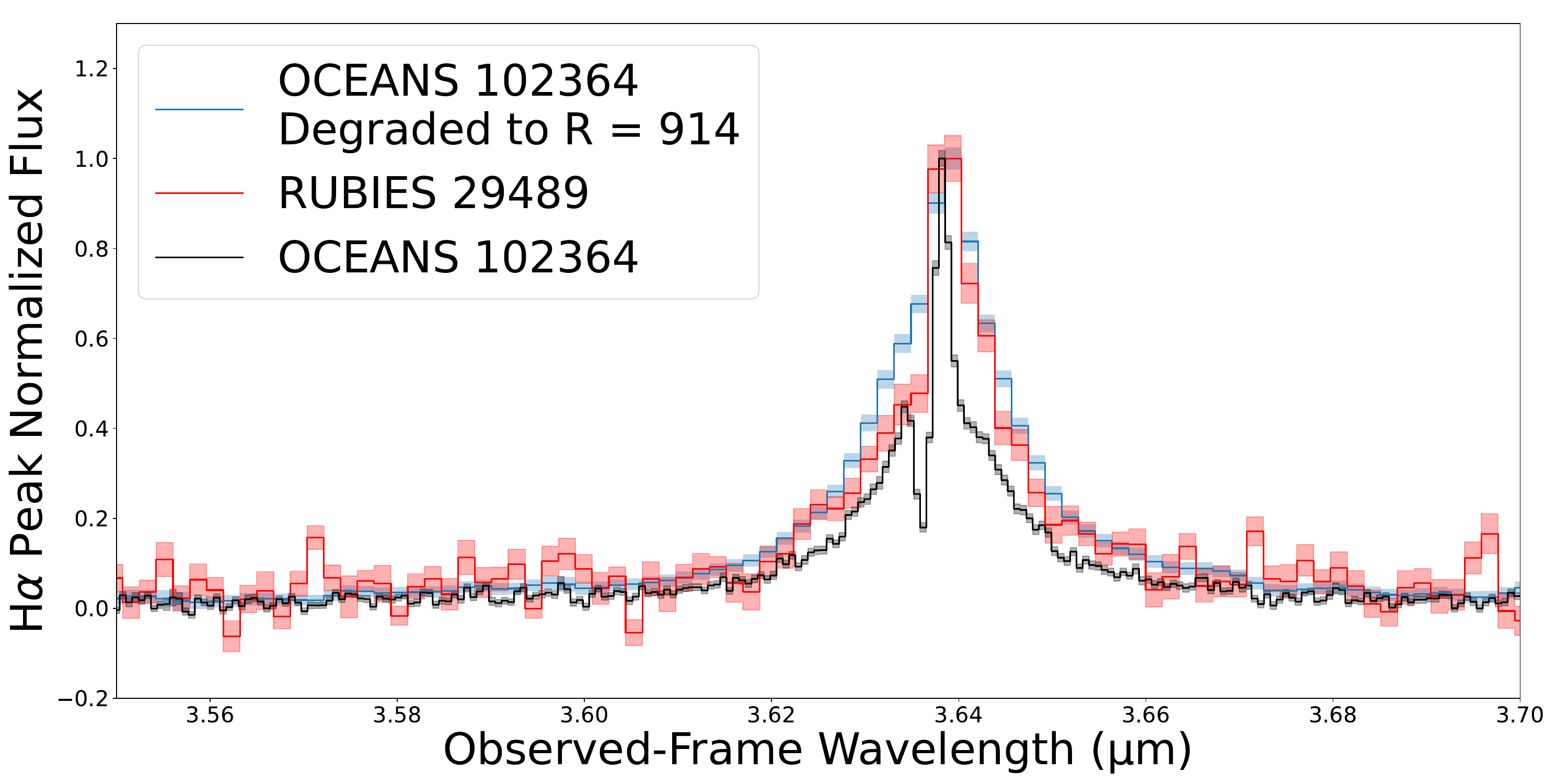}
    \caption{G395M (RUBIES) and G395H (OCEANS) coverage for a sources with an absorption feature only recovered in high-resolution spectroscopy. High-resolution OCEANS 102364 spectrum in black, medium-resolution RUBIES 28489 spectrum in red, and the OCEANS spectrum degraded to the RUBIES resolution for an \Ha\ line at $z=4.54$ in blue.} %\textbf{Bottom:} High-resolution OCEANS 101393 spectrum in black, medium-resolution RUBIES 37032 spectrum in red, and the OCEANS spectrum degraded to the RUBIES resolution for an \Ha\ line at $z=3.85$ in blue.  }
    \label{fig:res_effect}
\end{figure}

We identify 4 absorbing LRDs in OCEANS data, two of which were previously identified as absorbers (OCEANS 35829 in \citep{DEugenio2025, Lambrides2025} and OCEANS 20504 in \citet{Taylor2024}).
Figure \ref{fig:res_effect} shows an example of Balmer-absorption missed by previous spectroscopy. The target had low-resolution (prism) and medium-resolution (G395M) in RUBIES, and was not identified as a potential target with absorption in any previous studies. We plot the RUBIES G395M spectra in red, the OCEANS spectra in black, and the OCEANS spectra degraded to resolution that matches the RUBIES observation in blue. The absorption is too narrow ($\sigma_{abs}= 64 \,\rm{km}\, \rm{s}^{-1}$) to recover at G395M resolution. We note another absorber, OCEANS 100424, which similarly has narrow absorption ($\sigma_{abs}= 52 \,\rm{km}\, \rm{s}^{-1}$) that was not recoverable from a RUBIES observation of the source. %Potential future work may improve absorption recovery rates by defaulting to include absorption features in fitting, even in the case that they are not immediately apparent. 
%High-resolution spectroscopy was necessary to recover the absorption in the case of both sources. 
We note that OCEANS 102364, shown in Figure \ref{fig:res_effect}, is among the bluest LRDs from Figure \ref{fig:abscomps}. With an absorption line $\sim5$ pixels wide, the absorption would have been undetected at any lower resolution. %This source challenges ideas that Balmer absorption preferentially occurs in the reddest LRDs or those with the strongest optical slopes and may point to missed absorption in LRDs only recoverable at high spectral resolution.

\begin{figure}
    \centering
    \includegraphics[width=1\linewidth]{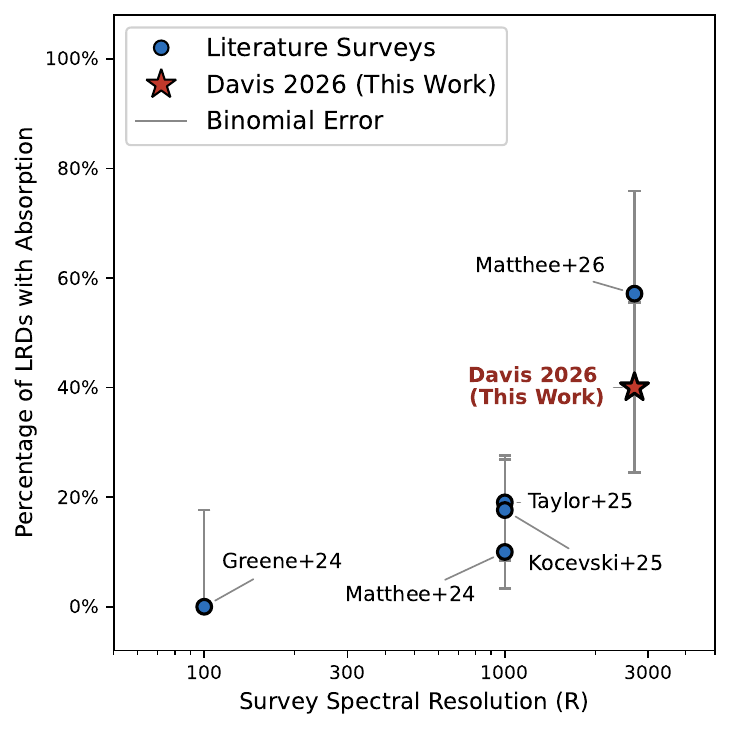}
    \caption{The fraction of LRDs with Balmer absorption in various surveys as a function of spectral resolution. At prism resolution ($R\sim100$), LRD surveys do not recover absorption (\citet{Greene2024}, 0/17). Of the surveys with medium-resolution ($R\sim1000$) coverage (\citet{Taylor2024} (4/21), \citet{Mathee2024} (2/20), and  \citet{Kocevski2024} (3/17)), typically 10-20\% of LRDs have Balmer line absorption. We note that \citet{Taylor2024} and \citet{Kocevski2024} utilized the same RUBIES dataset. In this work (4/10) and \citet{Matthee2026} (4/7), which both utilize high-resolution ($R\sim2700$) NIRSpec spectroscopy, absorption is recovered in $\sim$40-60\% of LRD targets and the samples do not overlap.}
    \label{fig:surveyfrac}
\end{figure}

Balmer-line absorption in LRDs was first identified in medium-resolution NIRspec spectroscopy from the RUBIES program, noted in several studies \citep{deGraaff2025, Hviding2025, Mathee2024, Taylor2024, Kocevski2024}. We review the detection fraction of LRD absorption in NIRSpec follow-up surveys contextualized by their survey resolution in Figure \ref{fig:surveyfrac}. Absorber percentages for other spectroscopic survey papers are plotted in blue and the fraction for this survey is plotted with a red star. We give the binomial standard error %rootp(1-p)/N) *100 to put it in percentage space
for the detection fractions and report a rule-of-three error ($3/\rm{N}$) for a null detection among $N$ sources. We also note that other observations of LRDs with deep, G395H spectroscopy have revealed Balmer absorption \citep{DEugenio2026, Kokorev2025, Torralba2025, Wang2025} and low-resolution spectroscopy has revealed absorption in the extreme ``cliff'' sources \citep{Naidu2025, Taylor2025} but we exclude single objects from this figure.

We include one study of LRDs that used low-resolution (R $\sim100$) NIRSpec prism spectroscopy \citep{Greene2024}. The work noted no Balmer-line absorption in the 17 LRDs surveyed. %We find only the extreme ``cliff'' sources, with Balmer breaks exceeding what is possible from stellar light, have absorption features recovered at low-resolution \citep{Naidu2025, Taylor2025}. 
At medium-resolution  (R $\sim1000$), surveys find Balmer-absorption in $10-20$\% of LRDs with \citet{Kocevski2024} finding 3 in 17, \citet{Mathee2024} finding 2 in 20, and \citet{Taylor2024} finding 4 in 21. At high-resolution (R $\sim2700$), absorption is detected in $50-60$\% of LRDs, with \citet{Matthee2026} finding 4 in 7 and this work finding 4 in 10. This climbing recovery rate with increased spectral resolution points to better recovery of absorption at higher spectral resolution. 

Such Balmer-absorption has been noted in $<1\%$ of Sloan Digital Sky Survey (SDSS) Quasars \citep{Leighly2025, Wang2015} but is observed with increased frequency in LRDs. Selection effects are not likely to impact the SDSS samples (R $= 2000$) and so Balmer-absorption must be uniquely common in LRD systems through some mechanism absent in most $z<1$ SDSS quasars. This points to Balmer-line absorption as a fundamental property of some LRDs. We further investigate Balmer absorption as a function of LRD properties in Section \ref{sec:propertiescomp}. 

\citet{Davis2026} showed that BLRs at $3<z<7$ are better recovered from medium-resolution NIRSpec spectroscopy when sources are targeted with increased exposure times (up from $\sim1$ hour to $\sim8$ hours) in the THRILS program. This approach successfully recovered 5 broad-line AGN originally targeted by the RUBIES program that were missed by previous searches for broad-line AGN within the RUBIES data \citep{Hviding2025, Taylor2024}. This approach, however, failed to recover any Balmer absorption in the targets with deep follow-up spectroscopy. Increased spectral resolution instead may be the driving observational effect that improves Balmer-line absorption recovery fractions since we do recover more absorption in the higher-resolution data. %although deeper observations would help constrain the low-luminosity LRD absorption fraction discussed in Section \ref{sec:lum}.

\subsubsection{Flux and Luminosity Selection Effect}
\label{sec:lum}

\begin{figure}
    \centering
    \includegraphics[width=1\linewidth]{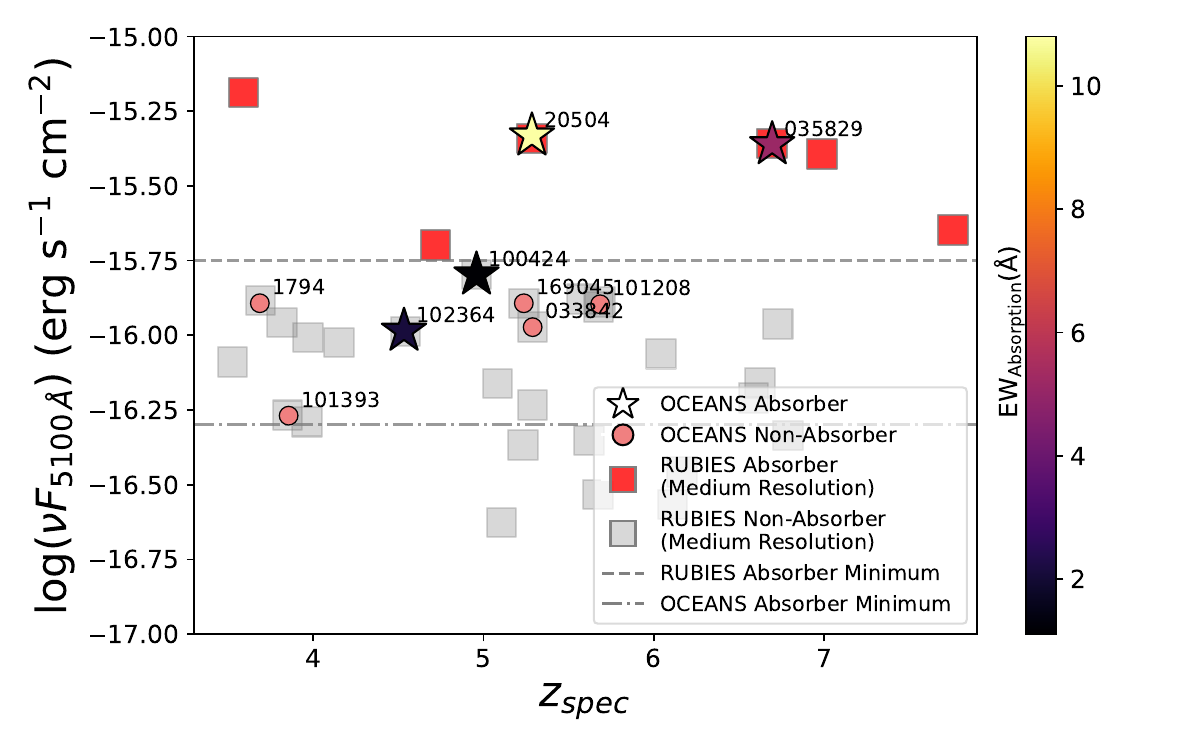}
    \includegraphics[width=1\linewidth]{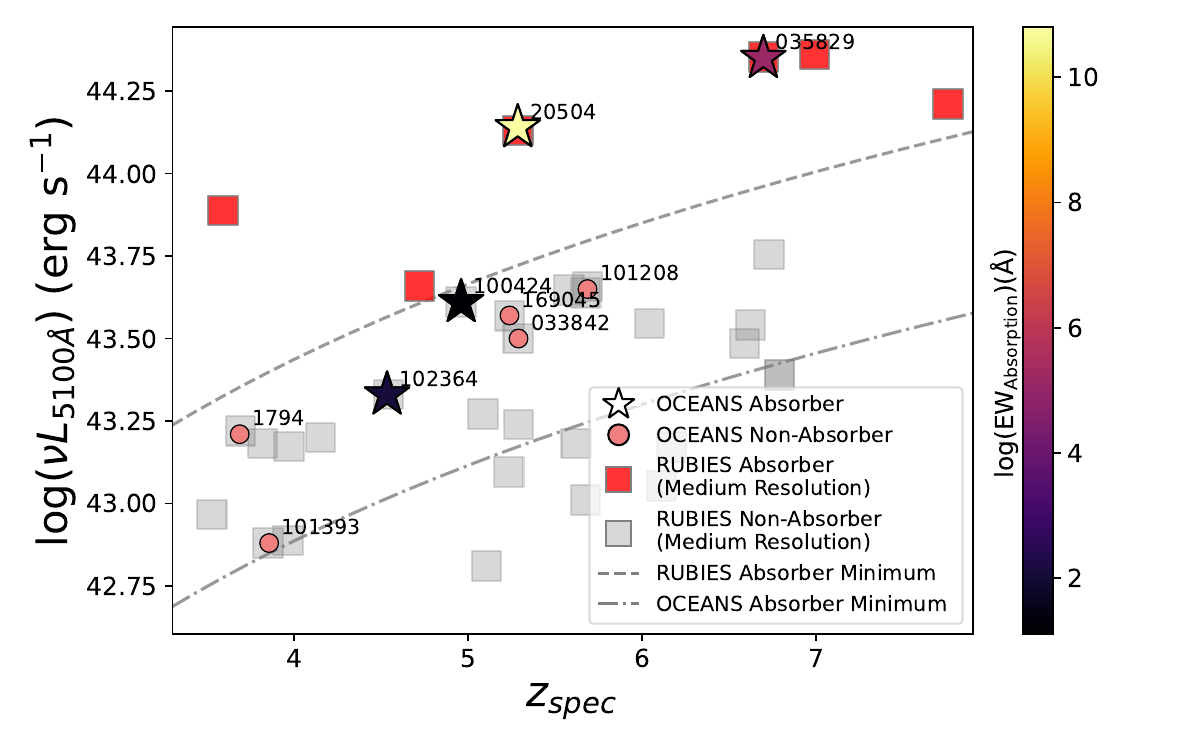}
    \caption{The luminosities and fluxes of LRDs observed by OCEANS and/or RUBIES programs in UDS and EGS. Squares indicate the medium-resolution RUBIES sample. Stars indicate OCEANS absorbers and circles indicate OCEANS LRDs without absorption. \textbf{Top:} Redshift vs $\nu F_{\nu, 5100} $ for both samples. We draw lines at the lowest flux absorber, dashed for RUBIES and dash-dotted  for OCEANS. \textbf{Bottom:} Redshift vs $\nu L_{\nu, 5100} $. We plot the flux-constraining lines from the top panel of this figure in Luminosity space.% An interesting version of this plot would have broad/narrow flux on x axis instead of zspec.
    }
    \label{fig:lumflux}
\end{figure}

We further investigate if LRD luminosity or flux influence absorber occurrence to probe whether LRDs are more likely to absorb if they are intrinsically brighter or if we observe absorption increasingly with detected flux. A correlation between absorption and flux might be associated with lower spectral signal-to-noise causing non-detection of absorption in fainter sources. We define both flux and luminosity at $5100$\AA\ which probes the rest-optical continuum, where the LRD ``engine''  is predicted to dominate the emission \citep{Kocevski2024}, while avoiding emission-line contamination. We measure the continuum from the low-resolution prism data. All OCEANS absorbers were previously observed by RUBIES and so can be directly compared.

%OCEANS sources without low-resolution data are excluded from the plots in this subsection because it is necessary to report the continuum flux confidently, especially for low-luminosity sources. 

Flux  measurements, $\nu F_\nu(5100)$, are plotted against the redshift distributions for OCEANS and RUBIES in the top panel of Figure \ref{fig:lumflux}. %We indicate the absorbing sources in RUBIES with red squares and the non-absorbers with gray squares. OCEANS sources are indicated by stars in the case that they are absorbers and coral circles in the case that they do not have detected absorption. OCEANS targets are colored according to their absorption EW. 
We draw horizontal lines for the dimmest detected absorber in both RUBIES and OCEANS. 
%Luminosity is an important LRD constraint for both completeness and covering fractions \citep{Lambrides2025}. 
Luminosity measurements, $\nu L_\nu(5100)$, are plotted for LRDs in the RUBIES + OCEANS spectroscopic samples in the bottom panel of Figure \ref{fig:lumflux}. We re-draw the dimmest detected absorber lines from the flux plot in luminosity space.

In the RUBIES data, the line tracing the dimmest source works as an effective separator, with all detected absorbers falling brighter than where the line lies in both plots. In the OCEANS data, the line is less of a clear delineator. The least-luminous absorbing OCEANS LRD is both the second least-luminous in the OCEANS sample and one of the least-luminous LRDs to be observed in either program. Furthermore, the LRDs between the RUBIES and OCEANS threshold lines are close to evenly split between absorbers and non-absorbers. %This points to OCEANS more efficiently probing the full spectrum of LRD luminosities and indicates a limit of 
%$\nu F_\nu 5100 \approx -15.75 ~\rm{erg}~\rm{s}^{-1}~\rm{cm}^{-2}$
%above which all LRDs with medium-resolution spectroscopy have absorption features. Such a limit is helpful for future spectroscopic targeting. 

We note that OCEANS 100424 and 102364 are among the least-luminous LRDs ever observed to have Balmer-line absorption. %Balmer absorption fraction as a function of luminosity is an important constraint on the covering factor of absorbing gas \citep{Lambrides2025}.
%Luminosity is an important LRD constraint for both completeness and covering fractions \citep{Lambrides2025} and so Balmer absorption fraction as a function of luminosity is an important constraint on the presence of in-flowing and out-flowing gas with covering fraction. %These sources represent a first probe into the low-luminosity absorbers. 
Both have bluer continuum slopes and optical colors (see Figure \ref{fig:abscomps}). These sources are a first look at absorption in low-luminosity LRDs and their typical absorption properties point to Balmer-line absorption as a more common phenomenon in LRDs than implied by medium-resolution spectroscopic data, which tends to recover only more luminous sources with broader absorption features.

\subsection{Balmer Absorption Properties}
\label{sec:propertiescomp}

% Balmer absorption arises from Hydrogen in the n=2 state at a narrow range of gas temperatures where such Hydrogen excitation is stable. This Hydrogen gas causes both the Balmer break observed in LRDs and the Balmer-line absorption identified here from high-resolution spectroscopy. Optical depth effects from cocoons of Hydrogen (see \citet{Naidu2025}) may drive strong breaks or density effects may contribute (see \citet{Madau2026, Tang2026, Ji2026}). We investigate trends from both the Balmer-line absorption and Balmer break as well as the optical to UV color ratios for the OCEANS spectroscopic sample. This is done in combination with the spectroscopic absorbers from \citep{Matthee2026} when possible, removing double counted absorbed in favor of the high-resolution OCEANS data.

\begin{figure}[t]
    \centering
    \includegraphics[width=1.1\linewidth]{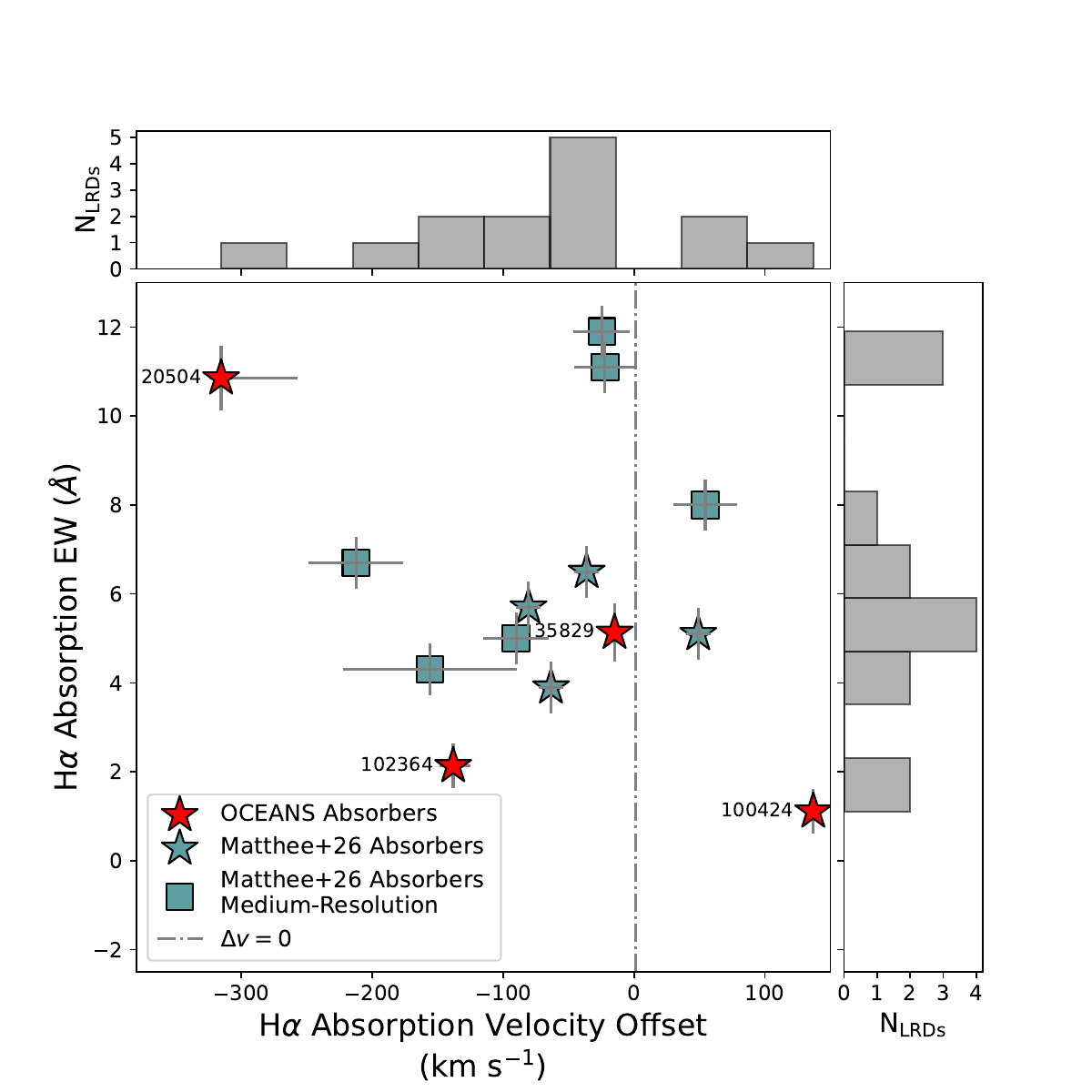}
    \caption{The velocity offset and rest frame EW of the \Ha\ absorption in the OCEANS (red) and \citep{Matthee2026} (blue) samples. High-resolution absorption detections (G395H and/or 235H) are marked with stars and medium-resolution (G395M) detections are marked with squares. We find a median EW of 5.3 \AA\ and a median velocity offset of $-49$~$\mathrm{km}~\mathrm{s}^{-1}$. Population distributions are given along each axis. }%for our OCEANS population and assign the absorbers from \citep{Matthee2026} the median errors from the OCEANS population. }
    \label{fig:allprops}
\end{figure}

\begin{figure*}
    \centering
    \includegraphics[width=1\linewidth]{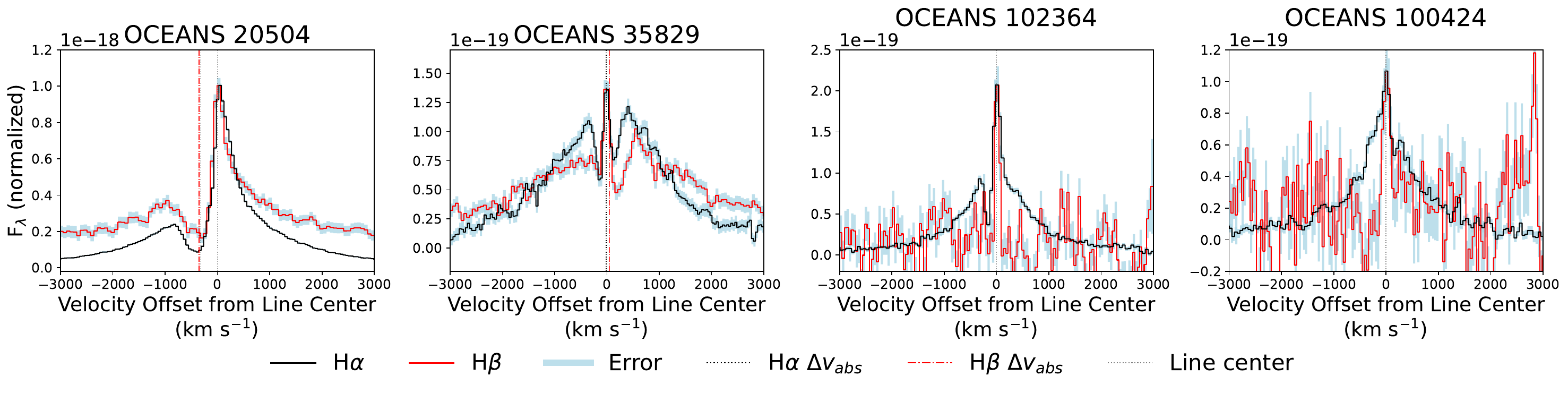}
    \caption{\Ha\ (black) and \Hb\ (red) line profiles for the OCEANS absorbing LRDs. \Hb\ is scaled to \Ha\ peak flux for profile comparison. Two absorbers have recovered absorption in their \Hb\ line profiles (left two) and two do not (right two) but are consistent with low signal-to-noise observations of \Hb. OCEANS 20504 and 35829 were both observed in multiple MSA pointings (see the Appendix) and the spectra displayed here are co-added. We note that the \Ha\ line in OCEANS 35829 fell in a chip gap for all but one of the MSA pointings that targeted it but \Hb\ was detected in all observations.}
    \label{fig:hahbprofiles} 
\end{figure*}

%We investigate all OCEANS absorbers through their absorption-line properties, including the sample of absorbing LRDs from \citet{Matthee2026} removing double counted absorbed in favor of the high-resolution OCEANS data.
We investigate the absorption-line properties of OCEANS LRDs. In this section, we include the sample of LRDs from \citet{Matthee2026}, removing two duplicate sources in favor of high-resolution OCEANS spectroscopy (OCEANS 20504, 35829). Figure \ref{fig:allprops} shows absorption EW and velocity offset for all LRDs including sources with high-resolution spectra (stars) and medium-resolution spectra (squares). Histograms along both axes indicate sample distributions. The population has a median absorption EW of 5.3 \AA\ and a median velocity offset of $-49$~$\mathrm{km}~\mathrm{s}^{-1}$. Figure \ref{fig:allprops} shows both EW and velocitty offset for all absorbers in the sample. The population has weak (EW$<12$ \AA) absorption and tends to be modestly blueshifted.
The majority (11/14 in the combined sample) of the absorption lines are blueshifted, consistent with outflowing gas that is viewed in the foreground of the LRD emission source. Four of these absorbers are best fit with Gaussians convolved with exponential wings (see the Appendix) implying either dense gas is scattering the photons \citep{Inayoshi2025, Rinaldi2025, Naidu2025} and/or the BLR is highly stratified \citep{Madau2026, Tang2026, Ji2026}.

\subsubsection{\Ha\ and \Hb\ Absorption Profile Comparison}
% \begin{figure}[h]
%     \centering
%     \includegraphics[width=1\linewidth]{offstall.pdf}
%     \caption{Velocity offset of all dual-absorber LRDs in the OCEANS sample. We mark each dual absorber with red points and plot a 1:1 agreement line in gray. We find that there is/is not evidence for variable absorption locations. (no for 20504, maybe for other after final pointing)}
%     \label{fig:offset}
% \end{figure}

\begin{figure*}
    \centering
    \includegraphics[width=.8\linewidth]{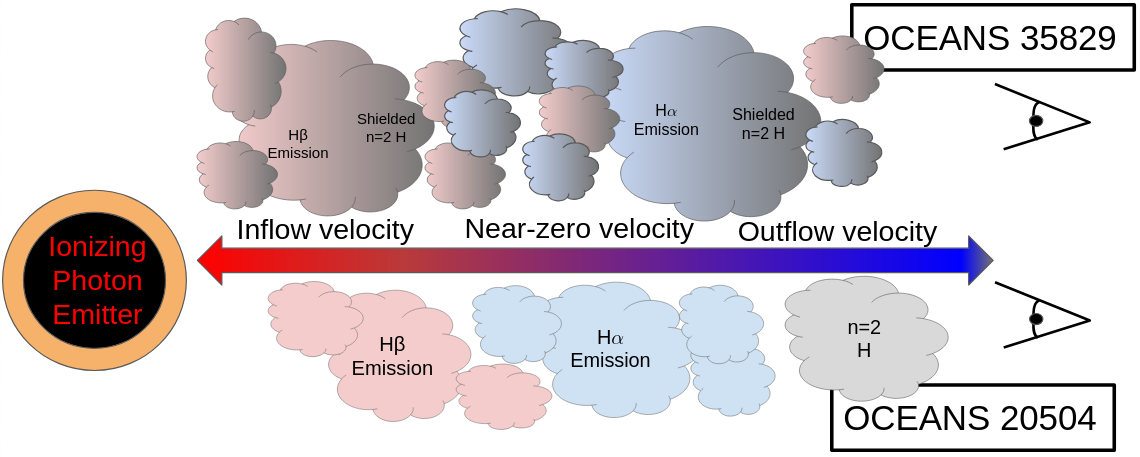}
    \caption{Cartoon of absorbing-medium locations for OCEANS 35829 and OCEANS 20504 which have absorption velocity offsets implying n=2 absorbing hydrogen is co-located with \Ha\ and \Hb\ emission regions which are radially stratified. This source contrasts the implied location of the absorbing medium location for OCEANS 20504 which has absorption velocity offsets implying n=2 absorbing hydrogen is located primarily outside of the emitting region. }
    \label{fig:oceans35829}
\end{figure*}

% Refer to Fig 8 here and note that the Ha and Hb profiles are fairly similar for the first 3 objects in terms of both emission and absorption features. The absorption features have the similar velocities and also have similar depths relative to the emission lines. The fourth object (102364) has low SNR in the Hb region and no significant broad Hb emission and so it is difficult to measure any Hb absorption.
%We can consider a plot of EW vs EW but it's probably ok to just discuss the similarity and point to Tables 3 and 4.
%Then discuss what it means for Ha and Hb to have similar absorption profiles in all of the LRDs for which it can be measured. 
Figure \ref{fig:hahbprofiles} shows the velocity offsets of \Ha\ and \Hb\ for the two dual-absorbing OCEANS LRDs (left two) and the line profiles for the other two absorbing OCEANS LRDs with only \Ha\ absorption (right two). \Ha\ profiles are plotted in black and \Hb\ in red. For dual absorbers, the absorption velocity offsets are indicated with vertical lines. 

In both OCEANS 20504 and 35829, the \Ha\ and \Hb\ profiles are similar. The absorption features share comparable profile shapes and depths relative to the emission lines with the exception of their velocity offsets which we discuss further here. The two absorbing LRDs that do not have \Hb\ absorption also do not have broad \Hb\ lines, consistent with the \Hb\ from the central engine being lost to low signal to noise. OCEANS 20504 and 35829 are the most luminous LRDs in the OCEANS sample and are the only OCEANS LRDs with dual absorption. This points to intrinsic luminosity or flux detection as a potential constraint for detecting absorption in multiple Balmer lines.

% Absorption has been observed in multiple Balmer lines for a handful of LRDs with deep spectroscopy. 
% One source, % THRILS 46403 (The ``GlimmIr'' \citep{Lambrides2025} /``Irony'' \citep{DEugenio2025}) , 
% OCEANS 35829, has been reported with different velocity offsets for both \Ha\ and \Hb, implying the absorption arises from different physical regions \citep{DEugenio2025,Lambrides2025}. Still another LRD shows absorption in HeI offset $\sim 150$ km/s from the Balmer-line absorption \citep{Juodzbalis2024}, albeit the velocity offsets of \Ha\ and \Hb\ absorption in this case agrees. Such line-dependent velocity offsets have been noted in Balmer emission for low-z quasars, attributed to  emission that is radially stratified and subject to different kinematics (see \citealt{Fries2024}). In contrast, Balmer-line absorption features typically share velocity offsets. %, but are less frequently identified in AGN \citep{Wang2015} and therefore difficult to constrain. 
% The only known AGN or LRD with offset Balmer absorption is OCEANS 35829, which has been studied previously at medium-resolution (R $\sim1000$). We provide a first look at high-resolution (R $\sim2700$) spectroscopic coverage for the source (OCEANS 35829) to compare EW ratios and velocity differences.

 %Another LRD shows absorption in HeI offset $\sim 150$ km/s from the Balmer-line absorption \citep{Juodzbalis2024}, albeit the velocity offsets of \Ha\ and \Hb\ absorption in this case agrees.  We provide a first look at high-resolution ($R \sim 2700$) spectroscopic coverage for the source.

The absorption in OCEANS 20504 has both velocity offsets and absorption EWs that are consistent between \Ha\ and \Hb\ (\Ha\ $\Delta v$ $=-315^{+58}_{-8.9} \rm{km \,} \rm{s}^{-1}$, \Hb\ $\Delta v$ $=-351^{+39}_{-35} \rm{km \,} \rm{s}^{-1}$, \Ha\ EW $=11^{+0.73}_{-0.27}$\AA, and \Hb\ EW $=6.0^{+0.64}_{-0.61}$\AA ). %\Ha\ absorption $\sim$1.5 times higher EW than \Hb. %This difference in EW is partly driven by the high system Balmer decrement (see Figure \ref{fig:balmdec}) and 
The large absorption velocity offset further implies the absorbing medium in OCEANS 20504 is spatially located in an outflowing region rather than interspersed within the emitting regions as demonstrated in OCEANS 35829. %We draw a schematic of the absorption medium location for OCEANS 20504 in Figure\ref{fig:oceans35829}.
Such agreement in velocity offset as well as absorption width (\Hb\ $\sigma_{abs}=266^{+27}_{-26} \rm{km \,} \rm{s}^{-1}$, \Ha\ $\sigma_{abs}=260^{+10}_{-7} \rm{km \,} \rm{s}^{-1}$)) implies that the absorption profiles have similar shapes for both lines. This consistency implies that the absorbing medium is located further from the system center than the Balmer emitting regions. 

OCEANS 35829, also called ``Irony'' \citep{DEugenio2025} and ``GlimmIr'' \citep{Lambrides2025}, has OCEANS Balmer absorption measurements of \Ha\ $\Delta v =-14^{+5.7}_{-6.4} \rm{km \,} \rm{s}^{-1}$ , \Hb\ $\Delta v = 51^{+26}_{-18} \rm{km \,} \rm{s}^{-1}$ , \Ha\ EW $=5.1^{+1.1}_{-0.66}$\AA, and \Hb\ EW $=3.7^{+1.4}_{-0.58}$\AA. We find that the EWs are statistically consistent between the two Balmer lines and measure a velocity offset between the absorption features of $65 \pm19\,\rm{km} \rm{s}^{-1}$, lower than the velocity offsets reported at medium resolution ($206 \pm 10 \rm{km \,} \rm{s}^{-1}$ in \citet{DEugenio2025} and $152 \pm 20 \rm{km \,} \rm{s}^{-1}$ in \citet{Lambrides2025}), but still a true velocity offset. Our values are in agreement with the \citet{Lambrides2025} measurement to $\sim3\sigma$ and the \citet{DEugenio2025} measurement to $\sim6 \sigma$, implying that absorption velocity offsets can be identified with medium-resolution NIRSpec spectroscopy but high-resolution is essential to constrain offsets between lines. %This greater disagreement with \citet{DEugenio2025} arises from the work's different fitting approach which fit multiple Gaussians to the absorber while the \citet{Lambrides2025} approach is  similar to this work. 

% Both \Ha\ and \Hb\ have absorption features that are near the systemic redshift center and so their absorption features are blended with both narrow and broad emission. This makes their absorption features degenerate with the strength of the narrow emission, which is more challenging to constrain the medium-resolution fit to the data. This highlights the need for high-resolution spectroscopy to resolve the full profile of the Balmer lines. At high-resolution, our fits to all line components are well constrained and unlikely to have velocity offsets driven by some resolution-dependent effect. 

Our OCEANS co-add has a median signal to noise per pixel of 11 around the \Hb\ emission line and a velocity width per pixel of $50 \, \rm{km} \, \rm{s}^{-1}$, allowing allows us to measure a line centroid to $\sim 1/2$ the pixel width ($\sim 25 \, \rm{km} \, \rm{s}^{-1}$). The \Ha\ offset of $-14 \rm{km} \, \rm{s}^{-1}$ is near-zero in velocity offset but the \Hb\ velocity offset, and therefore the reported difference in absorption velocity shift between \Ha\ and \Hb, is statistically significant.

% the +51 is assuredly a statistically significant positive offset given the uncertainties from the fit and the toy model above

% Rather than stating all of that, I think you can say that the S/N of the OCEANS data allow you to determine the line centers to ~1/2 the pixel width (based on the ~25 km/s error you quoted above)

% We note that the velocity offset is near the width of the spectral resolution. NIRSpec has a spectroscopic resolution element width of 2.2 detector pixels \citep{Jakobsen2022} which corresponds to a velocity offset of $\sim50 \,\rm{km \,} \rm{s}^{-1}$. However, the \Hb\ velocity offset is not consistent with 0 and so we report a statistically significant velocity offset between the absorption features accounting for instrumental resolution. 

%For the same reason, the absorption EWs are not directly comparable between medium and high resolution and so we focus our discussion on velocity offsets which are more robustly comparable.

% The strongest claim of an AGN or LRD with velocity offset Balmer absorption is OCEANS 35829, which has been studied previously at medium-resolution (R $\sim1000$), although velocity offsets between Balmer lines have had marginal detections in other studies \citep{DEugenio2025TWICE, DEugenio2026OTHER, Xihan2025}. We confirm the velocity offset in OCEANS 35829 between \Ha\ and \Hb\ in this work at $90 \pm15\,\rm{km} \rm{s}^{-1}$.
Balmer line-dependent velocity offsets have been noted in Balmer emission for low-$z$ quasars, attributed to emission that is radially stratified and subject to different kinematics (see \citealt{Fries2024, Villafana2024}). In contrast, Balmer-line absorption features typically share velocity offsets. Of note, one SDSS quasar with offset Balmer-line emission does not have offset Balmer-line absorption \citep{Wang2015}. Several SDSS identified quasars and other LRDs have detected Balmer absorption lines but their velocity offsets between Balmer lines are consistent \citep{Aoki2006, Hall2007, Juodzbalis2024, Leighly2025, Shi2016, Shi2017}.

We present another possibility; that the absorbing medium is co-located with the emitting regions in OCEANS 35829. Our cartoon model is sketched in Figure \ref{fig:oceans35829}. 
We turn to known properties of nearby AGN to explain the system.
Partial covering can produce optical depths that vary radially in BLRs, demonstrated in AGN along a single column density without invoking kinematically distinct absorbers \citep{Leighly2019, Leighly2025}. The absorption at different velocity offsets can be explained by the tendency of \Hb\ emission to occur closer to a SMBH than \Ha\ emission (evident by virial stratification of emission established in \citet{Korista2004}) with a radially distributed absorbing medium that has different optical depths driven by the covering fraction. In this framework, colder clumps of absorbing media where $n=2$ is thermally populated are co-spatially located with the Balmer emitting regions.

This may present as self-shielding clouds in the BLR where the side facing the emission source is subject to photoionization (with \Ha\ or \Hb\ emission ratios depending on proximity to the SMBH) while the side facing away from the ionizing photon emitting source is shielded and absorption dominates in this region. Such self-shielding clump structures exist in the torus around AGN \citep{Nenkova2008}. 
OCEANS 35829 may represent an LRD with a unique viewing angle, capturing two absorbing \Ha\ and \Hb\ dominated clouds co-located with the emission regions, that drives the observed velocity offsets.

Velocity offsets of \Ha\ and \Hb\ emission imply either a strong temperature or density gradient because \Ha\ and \Hb\ relative emission strengths are highly dependent on such gradients \citep{Inayoshi2025, Osterbrock1989}. 
However, absorption requires hydrogen in the $n=2$ state which is sensitive to the local physical conditions, with any single absorbing region subject to a single population state, making departures from standard velocity offsets more complicated than for Balmer emission. 

Balmer absorption can be sensitive to temperate and density gradients, for example in the photospheres of white dwarfs where Balmer lines have absorption velocity offsets of $\sim 10$\AA,  and extreme densities (log($n_e$) $\approx 17-18 \, \rm{cm^{-3}})$ \citep{Tremblay2009} drive Stark broadening. This is beyond what is realistic for LRD environments and direct line measurements from the medium-resolution data of this source imply densities of log$(n_e) \approx {4-5}$ \citep{DEugenio2025}, implying no single medium along the line-of-sight can create the observed velocity offsets. 
An alternative explanation requires an absorbing medium parameterized as distinct absorbers with non-standard optical depth weights with different velocity offsets driven by kinematic turbulent ``breathing modes'' which preferentially velocity shift \Hb\ absorption redward relative to \Ha\ absorption (see \citet{DEugenio2025} for a complete discussion of this model).

\begin{figure*}
    \centering
    \includegraphics[width=1\linewidth]{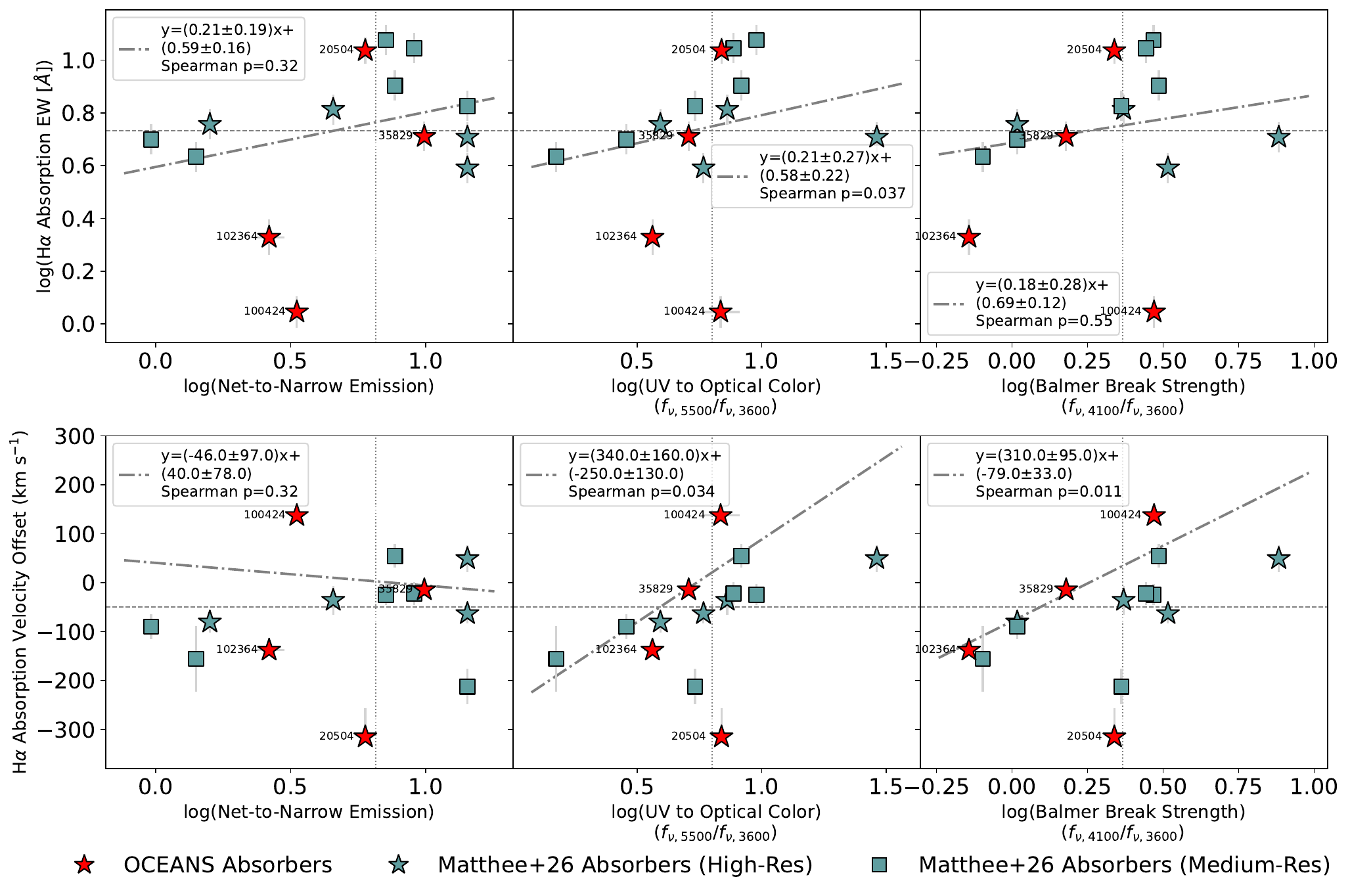}
    \caption{Absorption velocity offset from the systemic redshift (top row) and Absorption EW (bottom row) vs net-to-narrow emission (column 1),  UV-to-Optical color (column 2), and Balmer break strength (column 3) for the OCEANS and \citet{Matthee2026} absorbing LRDs. OCEANS data (G395H and G235H) is represented by red stars, high resolution absorbers (G395H and/or 235H) from \citep{Matthee2026} are represented by blue stars, and medium-resolution (G395M) absorbers from \citep{Matthee2026} are represented by blue squares.}%High-resolution absorption detections (G395H/235H) are marked with stars and medium-resolution (G395M) detections are marked with squares. }
    \label{fig:abscomps}
\end{figure*}

Dual-line absorption in LRDs provides a critical look at absorption velocity offsets that are difficult to measure from lower-resolution spectroscopy. While we recover only two dual-absorbers in OCEANS spectroscopy, the sources have distinct kinematically implied absorption structures. This could represent a viewing angle effect, where LRDs viewed from an angle capturing an outflow have $n=2$ hydrogen located primarily in the outflowing gas and LRDs viewed from an angle that do not capture strong outflows have $n=2$ hydrogen primarily co-located with emitting regions. Alternatively, OCEANS 35829 and 20504 could represent systems in different evolutionary phases, although they have similar UV to Optical colors and Balmer breaks in Figure \ref{fig:abscomps}, implying they have some systemic similarities. %Further, one or both of these systems could represent fringe cases. 
Future studies utilizing high-resolution NIRSpec spectroscopy will be essential to probe the kinematic nature of the absorbing gas in LRDs.
% We report covering fractions for the dual-absorbers (once the fits are finalized)

%\citep{Juodzbalis2024}

\subsubsection{Balmer Absorption Along the LRD Sequence}
\label{sec:popprop}

% plot 1-3: measure the absorprtion (Ew, velocity, width) vs LRD color 
% including no absorption

We investigate trends with color and absorption velocity offset first reported in \citet{Matthee2026} and look for other trends with Balmer break and net-to-narrow emission ratios in Figure \ref{fig:abscomps}. We include measurement uncertainties reported in the Appendix for the OCEANS absorbers and assign the median error from the OCEANS population to the \citet{Matthee2026} sources for EWs. Each dataset is fit by a weighted least-squares method and we report null probability $p$-values from Spearman correlation coefficients from \texttt{scipy} where a $p$-value $<0.05$ is a statistically significant trend. 

The leftmost panels in Figure \ref{fig:abscomps} compare absorption velocity offset and EW to net-to-narrow emission which we define as the ratio of total (broad$+$narrow$+$absorption) flux to the narrow-line flux. In both cases, we find no significant trend among the absorbers. %, with $p$ values of 0.32 for both EW and velocity offset. 
This net-to-narrow measurement is an efficient probe of LRD engine to host galaxy emission, with higher ratios implying the LRD central engine dominates over the host galaxy. A lack of correlation with either absorption property suggests that the dominance of the host galaxy has little to no influence on the absorption properties in LRDs. 

The middle panels of in Figure \ref{fig:abscomps} compare both velocity offset and EW to UV-to-optical color ($f_{\nu,5500}/f_{\nu,3600}$). The LRDs in the OCEANS sample have UV-to-optical flux ratios between 3 and 10, consistent with the center of the distribution of the \citet{Matthee2026} sources. We find significant correlations with both EW ($p=0.037$) and velocity offset ($p=0.034$). The trends imply that LRDs with redder colors (and therefore stronger optical continua with respect to the UV) have higher EW and outflows with redder velocities. %This color also contains the flux range utilized in Balmer break measurements and so this trend may reflect more available absorbing gas near the central engine. The correlation between absorption strength and the UV-to-optical color may also point to more hydrogen gas near the central engine driving stronger outflows. 
The UV-to-optical color traces the shape of the optical continuum which is highly sloped and best fit by a black-body radiative profile in many cases \citep{deGraaff2025BB}. This trend could also trace the shape, and therefore temperature, of the black body.

We also compare absorption properties directly with the strength of the Balmer break. The rightmost panels of Figure \ref{fig:abscomps} compare velocity offset and absorption EW to Balmer break strength ($f_{\nu,4100}/f_{\nu,3600}$). Balmer absorption arises from hydrogen in the $n=2$ state at a narrow range of gas states where Hydrogen excitation is stable. This same hydrogen gas drives the absorption causing the Balmer break observed in LRDs and so we investigate the kinematic offset of the $n=2$ hydrogen through interpreting the absorption features.  

The color used to define the Balmer break is similar to the color used to define the UV-to-optical ratio, and so exhibits similar trends, but probes the presence of $n=2$ hydrogen in the system more directly. We find a significant corelation with velocity offset ($p=0.011$) but not with EW ($p=0.55$). A lack of trend with EW could point to absorption strength being independent of available gas, but could also point to viewing angle and partial covering of the absorbing clouds, analogous to BALQSOs (e.g. \citet{Leighly2019, Leighly2025, Trump2006})
However, it may also imply that absorbing EW is highly model dependent and therefore difficult to constrain.

The positive trends between LRD color and absorption velocity are the strongest among any of the explored trends. This implies that LRDs with stronger Balmer breaks have absorbing gas with close to zero velocity or modestly redshifted while LRDs with less pronounced Balmer breaks have more $n=2$ hydrogen in outflows. This is consistent with our cartoon model in Figure \ref{fig:oceans35829}, where more available $n=2$ hydrogen implied by stronger Balmer breaks and redder colors means more available gas to drive the absorption. LRDs with modest absorption have $n=2$ gas primarily at larger radii while stronger breaks see absorbing gas included at smaller radii.

% This would imply available n=2 hydrogen dispersed throughout emission regions along the line of sight (consistent with near-zero velocity) with some LRDs having sufficiently abundant absorbing clumps near the SMBH such that they infall. 
In other words, LRDs with less available n=2 hydrogen tend to have that gas physically located further from the central SMBH in outflows, rather than co-located with emitters. This absorbing gas has a radial velocity gradient with closer gas preferring to infall while further gas outflows. Thus, redder LRDs tend to have absorption with near-zero or modestly red velocity offsets associated with more absorbing gas that is closer to the central engine. Such a model implies that Balmer break strength and UV-to-optical color are good tracers of the amount of absorbing gas along the line-of-sight. Absorption EW is sensitive to optical depth effects along the line-of-sight and carries less direct information about the amount of absorbing gas. This is further supported by our cartoon model which requires the absorbing gas be co-located with the broad emission gas and thus have a much lower covering fraction. 

This picture may describes OCEANS 100424, our only significantly redshifted \Ha\ absorber in OCEANS. The source has a Balmer break among the highest in the combined sample, and the highest in the OCEANS sample, but the smallest absorption EW. OCEANS 100424 may have most of this $n=2$ hydrogen near the system center, where it is infalling. Such deep $n=2$ hydrogen would be subject to a high line-of-sight optical depth and so we measure less Balmer-line absorption EW. 

OCEANS 20504 may represent another extreme of the outflow scenario. It has UV-to-optical colors near the middle of the population distribution and a Balmer break that is similarly near the median, but high-EW, high-velocity outflowing $n=2$ gas. This implies that some strong Balmer breaks can occur without the presence of shielded clumps, as in our cartoon of OCEANS 35829, but can instead also be driven by strong outflows along the line of sight, consistent with our cartoon model of this source. %OCEANS 20504 and another source from the \citet{Matthee2026} sample are the only sources in the population that lie in the center-bottom of the Balmer break vs velocity offset plot and may represent LRDs with n=2 hydrogen located in a physically different location. The n=2 hydrogen in these sources may have very high densities (comparable in break strength to sources like OCEANS 35829 with centrally located n=2 gas) that live primarily in the outflowing gas.

These two sources contrast the remaining blueshifted absorbing LRDs which follow the trend line, with the bluest velocity shifts tending to correlate with milder Balmer breaks. Such LRDs may have $n=2$ hydrogen located primarily in the outflowing gas (more consistent with the cartoon model of OCEANS 20504) that lies beyond the Balmer broad-line emitting regions. 
One such LRD, OCEANS 102364 is one of the bluest LRDs in the sample. Further, this represents one of the most blueshifted outflows among the LRDs and has the weakest Balmer break in the full sample. This implies outflows in an LRD with one of the weakest absorbing cocoons.

Balmer-line absorption in LRDs is a critical probe of the kinematics of hydrogen gas that drive the Balmer break spectral feature in LRDs. Although the number of Balmer-line absorbing LRDs is low, future high-resolution spectroscopic observations will expand this population and build a broader picture of the dynamic environments of excited hydrogen in LRDs.

\section{Summary and Future Work}
\label{sec:sum}

The OCEANS high-resolution spectroscopy program provides a new look at LRDs with Balmer absorption. The OCEANS sample was compiled from an amalgamation of targets in CEERS, and so we select LRDs that (1) satisfy the photometric criteria in \citet{Barro2026}, (2) are compact in rest-optical light, and (3) have broad \Ha\ and/or \Hb. We identify 10 LRDs from the OCEANS data, 4 of which have Balmer-line absorption.

Higher-resolution spectroscopy yields an increased recovery fraction for absorption features in LRDs, up to $40-60\%$ at $R\sim2700$ from $10-20\%$ at $R\sim1000$. This recovery is partly dependent on optical continuum flux for medium-resolution surveys. However, high-resolution surveys like OCEANS may probe the fraction of absorbers more efficiently as demonstrated through our recovery of absorption features lost in sources observed at lower spectral resolution. Absorption in LRDs tends to be near-zero in absorption velocity with a median velocity offset of $-49$ $\rm{km} \rm{s}^{-1}$ and median absorption EW of 5.3 \AA. 

OCEANS recovers two Balmer-absorbing LRDs with absorption in both \Ha\ and \Hb, both observed at high-resolution for the first time here. One source, OCEANS 20504 (also called RUBIES 42046) has absorption profiles consistent with a significant portion of the $n=2$ hydrogen occupying an outflow along the line of sight. The second dual-absorbing source, OCEANS 35829 (also called ``Irony'' \citep{DEugenio2025} and ``GlimmIr'' \citep{Lambrides2025}) has statistically significant velocity offsets for \Ha\ and \Hb\ absorption. The source is presented here in high spectral resolution for the first tine. We present a cartoon model that requires an absorbing medium co-located with \Ha\ and \Hb\ emission regions along the line of sight to explain the observation. 
We also note OCEANS 102364 as an LRD with Balmer absorption but a weak break and relatively blue UV-to-optical colors.

Trends relating absorption properties to LRD properties are explored to probe the nature of Balmer absorption along the LRD sequence. We report that LRD engine contribution compared to galaxy contribution to Balmer emission lines does not relate with observed absorption properties, implying the strengths of the LRD-powered broad emission lines have little correlation with the realtive galaxy strength. We also report and discuss trends relating absorption velocity offset to UV-to-optical colors and Balmer break strength and absorption EW to UV-to-optical color. Our strongest trends relate absorption velocity offset to Balmer break strength and UV-to-optical colors, implying LRDs with more available absorbing gas tend to have that gas more centrally located. 

% We note a difficulty in tying the absorption properties to other LRD features, especially due to a few outlier sources that do not follow clear population trends. However, we recover a positive corelation between absorber velocity offset and UV-to-optical color and a weak positive corelation between absorption EW and UV-to-optical color. There is no strong trend between absorption properties and narrow to net Balmer emission, implying that the emission dominance of the host galaxy does not influence the outflows from the central engine.

% We highlight OCEANS 20504 as an LRD with the highest absorption EW and bluest velocity offset of any absorbing LRD. This is contrasted by UV-to-optical colors near the median of the sample. OCEANS 102364 is also highlighted as one of the bluest known absorbing LRDs. We also note the recovery of Balmer-line absorption in two low-luminosity LRDs and one of the bluest reported LRDs with an absorption feature.
This work highlights the additional information gained by observing LRDs in high-resolution modes, giving a more comprehensive picture of the $n = 2$ hydrogen present in LRDs driving the strong Balmer breaks.
Future work on LRDs in CEERS and other fields will require additional high-resolution follow-up spectroscopy to build population distributions and probe the absorbing fraction fully across the LRD sequence.

% Future work: SPAM will recover Balmer breaks in all CEERS LRDs, future modeling efforts will need to account for dynamics from outflows/inflows

%\facility{HST (ACS, WFC3)}
\facility{JWST (NIRCam, NIRSpec)}

\software{\texttt{Astropy} \citep{astropy:2013, astropy:2018, astropy:2022} , \texttt{Matplotlib} \citep{Hunter2007}, \texttt{NumPy} \citep{numpy},
 \texttt{Pandas} \citep{pandas1, pandas2}, \texttt{Scipy} \citep{scipy}} %\citep{vanderwalt2011}

\begin{acknowledgments}

 We acknowledge the work of our colleagues in the CEERS, CAPERS, THRILS, and OCEANS collaborations and everyone involved in the JWST mission. We would like to thank Jackson White, Christopher Fontes, and Tan Hoang Bao Tran for their helpful opacity discussion. We would also like to thank the UMBRELA group at the Harvard \& Smithsonian Center for Astrophysics for the helpful dialogue.

 KD acknowledges support from a NSF Graduate Research Fellowship award number 2040433. MB acknowledges support from a NSF Graduate Research Fellowship.  KD, MB, RCS, and JRT acknowledge support from NASA JWST-GO-08410.018. KD acknowledges support from Los Alamos National Laboratory. %This work was released under LA-UR-xxx. 

\end{acknowledgments}

\bibliographystyle{aasjournal}
\bibliography{main}{}

%\newpage

\section{Appendix}
\label{sec:append}

We include additional information about our LRD population. Table \ref{tab:targets} lists the OCEANS IDs and gives matches to other surveys as well as target coordinates, OCEANS pointings, and lists the best-fit models for \Ha\ line profiles. The $\Delta \rm{BIC}$ fits to all possible profile shapes are shown in Figure\ref{fig:dbic}. Table \ref{tab:halpha}  give the \Ha\ line properties for the OCEANS LRDs inferred by their fits. Table \ref{tab:oiii} provides similar information for the \OIII\ fits and give systemic redshifts. %Figure \ref{fig:noabs} shows insets with the full line fits or the non-absorbing LRDs.

\begin{table*}[h]
\centering
\caption{LRD Target Descriptions}
\label{tab:targets}
\begin{tabular}{lccccl}
\hline\hline
OCEANS ID & Alternative ID & RA (deg) & Dec (deg) & Pointing  & Best Fit Model \\
\hline
020504 & RUBIES 42046 & 214.795369 & 52.788845 & 1, 4  & Exponential + Absorption \\
035829 & CAPERS 11585/THRILS(ID) & 214.892249 & 52.877403 & 2, 3, 6  & Gauss+ Abs \\
102364 & CAPERS 19300/RUBIES 29489 & 215.022070 & 52.920786 & 5  & Exp + Abs + Offset Gaussian \\
101393 & RUBIES 37032 & 214.849381 & 52.811828 & 4 & Exp + Abs \\
100424 & RUBIES 42232 & 214.886799 & 52.855377 & 3 & Gauss+ Abs \\
169045 & RUBIES 50052/CAPERS 11320/CEERS 2782 & 214.823454 & 52.830277 & 4 & Exp + Abs + Offset Gaussian \\
%\hline
033842 & RUBIES 60935 & 214.923373 & 52.925588 & 2  & Exp \\
101208 & RUBIES 37124 & 214.990980 & 52.916522 & 2, 5  & Exp + Offset Gaussian \\
%148165 & RUBIES 46985 & 214.805647 & 52.809502 & 1 & TBD  \\
%036603 & RUBIES 952625 & 214.975528 & 52.925265 & 5  & TBD\\
161695 & CEERS 1345 & 214.889684 & 52.832978 & 3, 6 &  Gauss \\
001794 & CAPERS 24745 & 214.814779 & 52.748950 & 1 & Exp + Offset Gaussian   \\

%021351 & CAPERS 50702 & 214.934950 & 52.844620 & 6 & -- & TBD (future obs)  \\
%159418 & CAPERS 56867 & 214.925600 & 52.851688 & 6 & -- & TBD (future obs)  \\

%051534 & CAPERS 86517 & 214.970770 & 52.920264 & 2 & -- & TBD NARROW LINE?COMPACT \\
\hline
% Dropped from sample &&&&&& Reason\\
% \hline
% %169018 & RUBIES 8488/CEERS 1019 & 215.035391 & 52.890672 & 5 & -- &  \OIII\ too broad \\
% 110484 & RUBIES 46724/CAPERS 92733 & 214.907747 & 52.882488 & 2 & -- & Ha outside of detector \\
%074500 & RUBIES 46423 & 214.828012 & 52.824095 & 4 & -- & featureless continuum \\
%036132 & CAPERS 12331 & 214.813212 & 52.817138 & 4 & -- & featureless continuum \\
%049342 & CAPERS 84296 & 214.790630 & 52.820811 & 4 & -- & featureless continuum \\
%094191 & CAPERS 19394 & 214.958408 & 52.875115 & 2 & -- & extended morphology \\
\end{tabular}
\end{table*}

\begin{table*}
\centering
\caption{H$\alpha$ emission and absorption-line properties. Absorption velocity offset
is measured relative to the systemic redshift.}
\label{tab:halpha}
\begin{tabular}{lcccccc}
\hline\hline
OCEANS ID &
$F_{\mathrm{narrow}}$ &
$F_{\mathrm{broad}}$ &
$F_{\mathrm{abs}}$ &
$\Delta v_{\mathrm{abs}}$ &
$\sigma_{\mathrm{abs}}$ &
EW$_{\mathrm{abs}}$ \\
 &
(erg s$^{-1}$ cm$^{-2}$) &
(erg s$^{-1}$ cm$^{-2}$) &
(erg s$^{-1}$ cm$^{-2}$) &
(km s$^{-1}$) &
(km s$^{-1}$) &
(\AA, rest) \\
\hline
%H ALPHA
20504 & $2.61 \pm 0.78 \times 10^{-17}$ & $1.50 \pm 0.04 \times 10^{-16}$ & $-3.06^{+0.13}_{-0.44} \times 10^{-17}$ & $-315^{+58}_{-9.0}$ & $260^{+10}_{-6.9}$ & $11^{+0.73}_{-0.27}$ \\
35829 & $5.95 \pm 5.94 \times 10^{-18}$ & $4.63 \pm 0.07 \times 10^{-17}$ & $-7.71^{+2.11}_{-5.88} \times 10^{-18}$ & $-14^{+5.7}_{-6.4}$ & $105^{+12}_{-9.8}$ & $5.1^{+1.1}_{-0.66}$ \\
102364 & $2.53 \pm 0.46 \times 10^{-18}$ & $1.80 \pm 0.03 \times 10^{-17}$ & $-1.40^{+0.26}_{-0.45} \times 10^{-18}$ & $-138^{+13}_{-10}$ & $64^{+7.0}_{-5.8}$ & $2.1^{+0.33}_{-0.24}$ \\
100424 & $4.30 \pm 0.43 \times 10^{-18}$ & $9.48 \pm 0.50 \times 10^{-18}$ & $-5.45^{+0.73}_{-0.78} \times 10^{-19}$ & $137^{+7.3}_{-6.9}$ & $52^{+6.2}_{-5.6}$ & $1.1^{+0.15}_{-0.14}$ \\

%non-absorbers
101393 & $3.90 \pm 16.40 \times 10^{-19}$ & $5.81 \pm 1.15 \times 10^{-18}$ & \nodata & \nodata & \nodata & \nodata \\
101208 & $7.70 \pm 1.21 \times 10^{-19}$ & $8.65 \pm 0.36 \times 10^{-18}$ & \nodata & \nodata & \nodata & \nodata \\
33842 & $1.90 \pm 0.19 \times 10^{-18}$ & $1.12 \pm 0.07 \times 10^{-17}$ & \nodata & \nodata & \nodata & \nodata \\
169045 & $1.05 \pm 0.01 \times 10^{-17}$ & $1.05 \pm 0.03 \times 10^{-17}$ & \nodata & \nodata & \nodata & \nodata \\
161695 & $7.63 \pm 0.81 \times 10^{-19}$ & $1.09 \pm 0.17 \times 10^{-18}$ & \nodata & \nodata & \nodata & \nodata \\
1794 & $8.58 \pm 0.18 \times 10^{-18}$ & $6.15 \pm 0.32 \times 10^{-18}$ & \nodata & \nodata & \nodata & \nodata \\
 &  &  &  &  &  &  \\
\hline
\end{tabular}
\end{table*}

\begin{table*}
\centering
\caption{H$\beta$ emission and absorption-line properties. Absorption velocity offset
is measured relative to the systemic redshift. 
}
\label{tab:hbeta}
\begin{tabular}{lcccccc}
\hline\hline
OCEANS ID &
$F_{\mathrm{narrow}}$ &
$F_{\mathrm{broad}}$ &
$F_{\mathrm{abs}}$ &
$\Delta v_{\mathrm{abs}}$ &
$\sigma_{\mathrm{abs}}$ &
EW$_{\mathrm{abs}}$ \\
 &
(erg s$^{-1}$ cm$^{-2}$) &
(erg s$^{-1}$ cm$^{-2}$) &
(erg s$^{-1}$ cm$^{-2}$) &
(km s$^{-1}$) &
(km s$^{-1}$) &
(\AA, rest) \\
\hline
 &  &  &  &  &  &  \\
20504 & $2.05 \pm 0.24 \times 10^{-18}$ & $1.14^{+0.06}_{-0.06} \times 10^{-17}$ & $-2.12^{+0.27}_{-0.30} \times 10^{-18}$ & $-35^{+38}_{-35}$ & $266^{+27}_{-26}$ & $6.0^{+0.64}_{-0.61}$ \\
%35829 & $6.65 \pm 1.57 \times 10^{-19}$ & $7.00^{+0.39}_{-0.16} \times 10^{-18}$ & $-9.54^{+1.00}_{-2.58} \times 10^{-19}$ & $75^{+20}_{-14}$ & $172^{+28}_{-11}$ & $4.4^{+0.93}_{-0.36}$ \\
35829 & $9.13 \pm 13.45 \times 10^{-19}$ & $8.17^{+0.27}_{-0.29} \times 10^{-18}$ & $-1.11^{+0.24}_{-1.31} \times 10^{-18}$ & $51^{+26}_{-18}$ & $153^{+23}_{-30}$ & $3.7^{+1.4}_{-0.58}$ \\
102364 & $6.10 \pm 0.82 \times 10^{-19}$ & \nodata & \nodata & \nodata & \nodata & \nodata \\
100424 & $4.66 \pm 1.43 \times 10^{-19}$ & \nodata & \nodata & \nodata & \nodata & \nodata \\
%absorbers
101393 & \nodata & \nodata & \nodata & \nodata & \nodata & \nodata \\
%101208 & $1.86 \pm 1.34 \times 10^{-19}$ & $6.76^{+2.23}_{-1.75} \times 10^{-19}$ & \nodata & \nodata & \nodata & \nodata \\
101208 & $1.86 \pm 1.34 \times 10^{-19}$ & \nodata & \nodata & \nodata & \nodata & \nodata \\
33842 & $5.57 \pm 1.89 \times 10^{-19}$ & \nodata & \nodata & \nodata & \nodata & \nodata \\
169045 & $3.11 \pm 0.10 \times 10^{-18}$ & $1.98^{+0.29}_{-0.28} \times 10^{-18}$ & \nodata & \nodata & \nodata & \nodata \\
161695 & $2.31 \pm 0.92 \times 10^{-19}$ & $4.09^{+1.74}_{-1.63} \times 10^{-19}$ & \nodata & \nodata & \nodata & \nodata \\
1794 & $3.35 \pm 0.07 \times 10^{-18}$ & \nodata & \nodata & \nodata & \nodata & \nodata \\
%non-absorbers

 &  &  &  &  &  &  \\
\hline
\end{tabular}
\end{table*}

\begin{table*}
\centering
\caption{[O\,{\sc iii}]$\lambda\lambda$4960,5008 emission-line properties. OCEANS 101393 does not have \OIII\ line coverage and so we use the narrow \Ha\ peak to calculate a systemic redshift.
}
\label{tab:oiii}
\begin{tabular}{lccccc}
\hline\hline
OCEANS ID &
$F_{\mathrm{narrow}}^{ [\rm{OIII}] 5008}$ &
$F_{\mathrm{broad}}^{[\rm{OIII}] 5008}$ &
$F_{\mathrm{narrow}}^{[\rm{OIII}] 4960}$ &
$F_{\mathrm{broad}}^{[\rm{OIII}] 4960}$ &
$z_{[\rm{OIII}]5008 }$ \\
 &
(erg s$^{-1}$ cm$^{-2}$) &
(erg s$^{-1}$ cm$^{-2}$) &
(erg s$^{-1}$ cm$^{-2}$) &
(erg s$^{-1}$ cm$^{-2}$) &
 \\
\hline

 &  &  &  &  &  \\
20504 & $7.74^{+0.14}_{-0.13} \times 10^{-18}$ & $2.01^{+0.14}_{-0.15} \times 10^{-18}$ & $2.57^{+0.04}_{-0.04} \times 10^{-18}$ & $6.68^{+0.48}_{-0.49} \times 10^{-19}$ & 5.276 \\
35829 & $6.98^{+0.07}_{-0.07} \times 10^{-18}$ & $8.24^{+1.09}_{-1.10} \times 10^{-19}$ & $2.32^{+0.02}_{-0.02} \times 10^{-18}$ & $2.74^{+0.36}_{-0.37} \times 10^{-19}$ & 6.684 \\
102364 & $2.69^{+0.07}_{-0.07} \times 10^{-18}$ & \nodata & $8.94^{+0.24}_{-0.24} \times 10^{-19}$ & \nodata & 4.542 \\
100424 & $1.30^{+0.07}_{-0.07} \times 10^{-18}$ & \nodata & $4.31^{+0.23}_{-0.22} \times 10^{-19}$ & \nodata & 4.953 \\

%non-absorbers
101393 & $1.15^{+0.05}_{-0.05} \times 10^{-18}$ & \nodata & $3.84^{+0.16}_{-0.16} \times 10^{-19}$ & \nodata & 3.850 \\
101208 & $9.84^{+0.33}_{-0.33} \times 10^{-19}$ & \nodata & $3.27^{+0.11}_{-0.11} \times 10^{-19}$ & \nodata & 5.682 \\
33842 & $3.09^{+0.12}_{-0.12} \times 10^{-18}$ & \nodata & $1.03^{+0.04}_{-0.04} \times 10^{-18}$ & \nodata & 5.287 \\
169045 & $1.64^{+0.01}_{-0.01} \times 10^{-17}$ & $3.08^{+0.12}_{-0.12} \times 10^{-18}$ & $5.45^{+0.04}_{-0.04} \times 10^{-18}$ & $1.02^{+0.04}_{-0.04} \times 10^{-18}$ & 5.239 \\
161695 & $8.92^{+0.34}_{-0.35} \times 10^{-19}$ & \nodata & $2.97^{+0.11}_{-0.12} \times 10^{-19}$ & \nodata& 5.666 \\
1794 & $1.33^{+0.01}_{-0.01} \times 10^{-17}$ & \nodata & $4.41^{+0.03}_{-0.03} \times 10^{-18}$ & \nodata & 3.681 \\
 &  &  &  &  &  \\
\hline
\end{tabular}
\end{table*}

\end{document}